\documentclass[11pt,preprint,twocolumn]{aastex631}

\usepackage{amsmath}
\usepackage{threeparttable}
\usepackage{amsmath}	
\usepackage{amssymb}	
\usepackage{cases}
\usepackage{color}
\usepackage{hyperref}
\usepackage{xcolor}
\usepackage{multirow}
\usepackage{array}
\usepackage{amsmath,amstext}
\usepackage{booktabs}
\usepackage{cleveref}
\usepackage{paralist}
\usepackage{graphicx}
\graphicspath{{Figure/}}

\def\max{{\rm max}}

\newcommand{\teff}{$T_{\mathrm{eff}}$}
\newcommand{\logg}{$\log{g}$}
\newcommand{\feh}{$[\mathrm{Fe/H}]$}
\newcommand{\alp}{$[\alpha/\mathrm{M}]$}

\usepackage{xcolor}
\definecolor{orcidgreen}{RGB}{166, 206, 57}
\usepackage{hyperref}

\shorttitle{Stellar Parameters of M dwarfs }
\shortauthors{Qiu et al.}

\graphicspath{{./}{Fig/}}

\begin{document}

\title{Stellar Parameters of BOSS M dwarfs in SDSS-V DR19}

\author[0000-0002-8280-4808]{Dan Qiu}
\affiliation{Key Laboratory of Space Astronomy and Technology, National Astronomical Observatories, CAS, Beijing 100101, People’s Republic of China}\email{E-mail: dqiu1223@gmail.com; liuchao@nao.cas.cn}
\affiliation{University of Chinese Academy of Sciences, Beijing 100049, People’s Republic of China}
\affiliation{Department of Astronomy, The Ohio State University, 140 W. 18th Ave., Columbus, OH 43210, US}

\author{Jennifer A. Johnson}
\affiliation{Department of Astronomy, The Ohio State University, 140 W. 18th Ave., Columbus, OH 43210, US}

\author{Chao Liu}
\affiliation{Key Laboratory of Space Astronomy and Technology, National Astronomical Observatories, CAS, Beijing 100101, People’s Republic of China}
\affiliation{University of Chinese Academy of Sciences, Beijing 100049, People’s Republic of China}
\affiliation{Institute for Frontiers in Astronomy and Astrophysics, Beijing Normal University, Beijing 100875, People’s Republic of China}
\affiliation{Zhejiang Lab, Hangzhou, Zhejiang 311121, People’s Republic of China}

\author[0000-0002-7883-5425]{Diogo Souto}
\affiliation{Departamento de F\'isica, Universidade Federal de Sergipe, Av. Marcelo Deda Chagas, S/N Cep 49.107-230, S\~ao Crist\'ov\~ao, SE, Brazil}

\author[0000-0003-3410-5794]{Ilija Medan}
\affiliation{Department of Physics and Astronomy,Vanderbilt University,
Nashville, TN 37235, USA}

\author[0000-0003-1479-3059]{Guy S. Stringfellow}
\affiliation{Center for Astrophysics and Space Astronomy, 
Department of Astrophysical and Planetary Sciences, 
University of Colorado 389-UCB, Boulder, CO 80309, US}

\author[0000-0003-0179-9662]{Zachary Way}
\affiliation{Department of Physics and Astronomy, Georgia State University, 25 Park Place, Atlanta, GA 30303, USA}

\author{Yuan-sen Ting}
\affiliation{Department of Astronomy, The Ohio State University, 140 W. 18th Ave., Columbus, OH 43210, US}

\author[0000-0003-0174-0564]{Andrew R. Casey}
\affiliation{School of Physics \& Astronomy, Monash University, Australia}
\affiliation{Center for Computational Astrophysics, Flatiron Institute, 162 Fifth Avenue, New York, NY 10010, USA}

\author[0000-0002-0149-1302]{B\'arbara Rojas-Ayala}
\affiliation{Instituto de Alta Investigaci\'on, Universidad de Tarapac\'a, Casilla 7D, Arica, Chile}

\author[0000-0002-7795-0018]{Ricardo L\'opez-Valdivia}
\affiliation{Instituto de Astronom\'ia, Universidad Nacional Aut\'onoma de M\'exico, Ap. 106,  Ensenada 22800, BC, M\'exico}

\author[0000-0002-6270-8851]{Ying-Yi Song}
\affiliation{David A. Dunlap Department of Astronomy \& Astrophysics, University of Toronto, 50 St. George Street, Toronto, ON M5S 3H4, Canada}
\affiliation{Dunlap Institute for Astronomy \& Astrophysics, University of Toronto, 50 St. George Street, Toronto, ON M5S 3H4, Canada}

\author[0000-0002-6434-7201]{Bo Zhang}
\affiliation{Key Laboratory of Space Astronomy and Technology, National Astronomical Observatories, CAS, Beijing 100101, People’s Republic of China}

\author[0000-0002-3651-5482]{Jiadong Li}
\affiliation{Max-Planck-Institut für Astronomie, Königstuhl 17, D-69117 Heidelberg, Germany}

\author[0000-0003-0012-9093]{Aida Behmard}
\affiliation{Center for Computational Astrophysics, Flatiron Institute, 162 Fifth Ave, New York, NY 10010, USA}

\author[0000-0001-8237-5209]{Szabolcs~M{\'e}sz{\'a}ros}
\affiliation{ELTE E\"otv\"os Lor\'and University, Gothard Astrophysical Observatory, 9700 Szombathely, Szent Imre H. st. 112, Hungary}
\affiliation{MTA-ELTE Lend{\"u}let "Momentum" Milky Way Research Group, Hungary}

\author[0000-0002-3481-9052]{Keivan G.\ Stassun}
\affiliation{Department of Physics and Astronomy, Vanderbilt University, Nashville, TN 37235, USA}

\author{Jos\'e G. Fern\'andez-Trincado}
\affiliation{ Universidad Cat\'olica del Norte, N\'ucleo UCN en Arqueolog\'ia Gal\'actica - Inst. de Astronom\'ia, Av. Angamos 0610, Antofagasta, Chile}

\begin{abstract}

We utilized the Stellar LAbel Machine (SLAM), a data-driven model based on Support Vector Regression, to derive stellar parameters (\feh, \teff, and \logg) for SDSS-V M dwarfs using low-resolution optical spectra (R$\sim$2000) obtained with the BOSS spectrographs. These parameters are calibrated using LAMOST F, G or K dwarf companions (\feh), and APOGEE Net (\teff\ and \logg), respectively. 
Comparisons of SLAM predicted \feh\ values between two components of M+M dwarfs wide binaries show no bias but with a scatter of 0.11 dex. Further comparisons with two other works, which also calibrated the \feh\ of M dwarfs by using the F/G/K companions, reveal biases of -0.06$\pm$0.16 dex
and 0.02$\pm$0.14 dex, respectively. The SLAM-derived effective temperatures agree well with the temperature which is calibrated by using interferometric angular diameters (bias: -27$\pm$92 K) and those of the LAMOST (bias: -34$\pm$65 K), but are systematically lower than those from an empirical relationship between the color index and \teff\ by 146$\pm$45 K. The SLAM surface gravity aligns well with those of LAMOST (bias: -0.01$\pm$0.07 dex) and those derived from the stellar mass and radius (bias: -0.04$\pm$0.09 dex).  Finally, we investigated a bias in \feh\ between SLAM and APOGEE ASPCAP. It depends on ASPCAP's \feh\ and \teff, we provide an equation to correct the ASPCAP metallicities. 

\end{abstract}

\keywords{methods: statistical -- stars: evolution, fundamental parameters, low-mass -- Galaxy: stellar content}

\section{Introduction}\label{sect:intro}
The fifth generation of the Sloan Digital Sky Survey 
\citep[SDSS-V;][]{Kollmeier-2025} comprises three primary scientific programs: 
the Milky Way Mapper \citep[MWM;][]{M-2024eas..conf1087M}, the Black Hole Mapper \citep[BHM;][]{Anderson-2023}, and the Local Volume Mapper \citep[LVM;][]{Drory-2024AJ....168..198D}. 
Among these, the MWM is specifically designed to unravel the Milky Way's structure, composition, dynamics, and evolutionary history through detailed spectroscopic observations of its stellar populations.

Determining accurate stellar parameters is crucial for advancing our understanding of stellar astrophysics and the evolution of galaxies. Key parameters such as effective temperature (\teff), surface gravity (\logg), and metallicity (\feh) provide essential insights into the physical properties, chemical composition, and evolutionary stages of stars. These parameters serve as foundational data for studying stellar populations, modeling stellar atmospheres, and interpreting galactic formation and evolution \citep[e.g.,][]{Nissen-1997A&A...326..751N, Casagrande-2011A&A...530A.138C,Jofr-2019}. Precise measurements of these properties also enable the identification of unique stellar phenomena, such as chemically peculiar stars, and support the characterization of exoplanetary systems by linking stellar properties to planetary formation processes \citep{Adibekyan-2012A&A...545A..32A, Sousa-2006A&A...458..873S}. Therefore, a reliable and consistent determination of stellar parameters is indispensable for both theoretical and observational astronomy.

In SDSS-V, MWM is using the APOGEE spectrographs 
\citep{Wilson-2019PASP..131e5001W} and the BOSS spectrographs \citep{Smee-2013AJ....146...32S} to observe stars throughout the HR diagram. These spectrographs can observe the whole sky, with one set on the Sloan Foundation Telescope \citep{Gunn-2006AJ....131.2332G} at Apache Point Observatory (APO) in New Mexico and another at the Irenee Du Pont telescope \citep{Bowen-1973ApOpt..12.1430B} at Las Campanas Observatory (LCO) in Chile. The APOGEE spectrographs have 300 fibers, a spectral resolution of  R$\sim$22,500 and cover the near-infrared H-band. MWM is targeting infrared bright (H $<$ 13) stars, particularly luminous red giants, YSOs, planet hosts, and solar neighborhood stars. The APOGEE spectra are reduced using an updated version of the APOGEE Data Reduction Pipeline \citep[][Nidever et al. in prep]{Nidever-2015AJ....150..173N}. With the BOSS spectrographs, MWM is targeting M dwarfs, white dwarfs, compact binaries, and YSOs to G$\lesssim$20. The BOSS spectrographs have 500 fibers and cover the optical range (3800-9800 \AA) at lower resolution (R$\sim$2000). These spectra are reduced by the BOSS Data Reduction Pipeline (Morrison et al., in prep). The 1-D wavelength-calibrated spectra are then passed to the data analysis pipelines for stellar parameter and abundance determination via Astra (Casey et al. in prep, v\_0.6.0). 

Astra is a software package designed to manage and coordinate the analysis of spectra from the MWM program. The development of Astra is motivated by the expanded scope and diversity of stellar targets in SDSS-V compared to earlier SDSS phases. 
SDSS-V MWM encompasses a broader range of stellar types and evolutionary stages, from M dwarfs to white dwarfs. The varying physical conditions in the photospheres of these stars necessitate flexible and adaptable analysis pipelines, as methods suitable for one stellar type may not be applicable to another. Astra addresses this challenge by integrating multiple analysis pipelines into a unified framework, ensuring self-documented and scientifically robust data products.
It processes
reduced data products using several data analysis pipelines. Of relevance to this paper, APOGEE spectra are analyzed with the ASPCAP \citep[][M{\'e}sz{\'a}ros et al. in prep]{Garc-2016AJ....151..144G}, and Apogee Net \citep{Olney-2020AJ....159..182O,Sprague-2022AJ....163..152S,Sizemore-2024AJ....167..173S} pipelines. The BOSS spectra analyzed with BOSS Net \citep{Sizemore-2024} and SLAM (this paper) data reduction pipelines. The output of Astra are science-ready data products that include stellar parameters, chemical abundances, line measurements, and other critical quantities.

M dwarfs are the smallest and coolest stars on the main sequence. As the most abundant stellar type in the Milky Way, comprising approximately 70\% of all stars\citep{Henry-2006AJ....132.2360H,Bochanski-2010}, they serve as essential tracers of the Galaxy's chemical and dynamical history \citep[e.g.][]{Bochanski-2007AJ....134.2418B,Chabrier-2003PASP..115..763C,Winters-2019}. Furthermore, their atmospheres retain signatures of the environments in which they formed, and their long lifespans make them invaluable for studies of Galactic archaeology \citep[e.g.][]{Freeman-2002ARA&A..40..487F,West-2011, Woolf-2012,Allard-2012RSPTA.370.2765A}. M dwarfs also play a key role in exoplanet research, as many potentially habitable planets have been detected orbiting these stars, leveraging their small radii and low luminosities to enhance planet detectability \citep{Mann-2011,Gaido-2016MNRAS.457.2877G,Ribas-2018}.

The SDSS-V offers an unprecedented opportunity to study M dwarfs on a massive scale. And a critical step in achieving the above goals is the precise determination of stellar atmospheric parameters, such as \teff, \logg, \feh, and chemical abundances. These parameters are fundamental for characterizing M-type stars and interpreting their role in the galaxy's history \citep[e.g.][]{Nissen-2010A&A...511L..10N,Rix-2013A&ARv..21...61R,Ting-2019ApJ...879...69T}. However, the intrinsic complexity of M dwarfs—the result of their cool atmospheres, molecular absorption features, and sensitivity to metallicity—poses unique challenges for parameter estimation \citep{Allard-2012RSPTA.370.2765A,Mann-2013a,Mann-2015}, especially for metallicity. For decades, the community has relied on synthetic spectra from 1D hydrostatic model atmospheres like PHOENIX, BT-Settl, and MARCS \citep{Allard-2012RSPTA.370.2765A,Baraffe-2015,Allard-2016a,Van-2017A&A...601A..10V}. However, these models suffer from significant systematic uncertainties stemming from incomplete molecular and atomic line lists, missing opacity sources, and the simplified assumption of local thermodynamic equilibrium (LTE), which often breaks down in the upper layers of these cool atmospheres \citep{Rajpurohit-2018A&A...620A.180R,Jofr-2019}. These model-based inaccuracies are particularly severe for metallicity, an essential tracer of chemical evolution, as changes in metal content intricately affect the entire spectral energy distribution by altering atmospheric opacity and molecular formation efficiencies \citep{Veyette-2017}.

To circumvent these modeling limitations, significant progress over the past decade has come from high-resolution spectroscopic surveys operating in the near-infrared (NIR), such as APOGEE \citep{Abdurro-2022ApJS..259...35A} and CARMENES \citep{Quirrenbach-2016SPIE.9908E..12Q}. The NIR is advantageous because molecular line density is lower and the continuum is better defined, allowing for the use of more isolated atomic lines that are better modeled and serve as more reliable abundance indicators \citep{Souto-2020ApJ...890..133S,Passegger-2019}. These high-resolution studies have been crucial for creating benchmark samples of M dwarfs with precisely determined parameters, forming the foundation of our modern understanding. However, these surveys are generally limited to smaller sample sizes compared to massive optical surveys. Consequently, determining precise and accurate metallicities for the millions of M dwarfs observed in low-resolution optical spectra remains a major unsolved problem. This is because the optical region is heavily blanketed by overlapping molecular features that are highly sensitive to atmospheric parameters, making it extremely difficult for theoretical models to disentangle the effects of \teff, \logg, and \feh.

In recent years, data-driven methods, such as The Payne/DD-Payne \citep{Ting-2019ApJ...879...69T,Xiang-2019ApJS..245...34X} and the Stellar LAbel Machine \citep[SLAM;][]{Zhang-2020}, have emerged as powerful alternatives. These approaches leverage large spectral datasets and benchmark training labels to bypass the direct reliance on imperfect physical models, enabling the prediction of stellar parameters with high precision for large datasets. A major limitation to apply data-driven methods is the lack of benchmark M dwarfs with well-determined metallicity. Fortunately, wide binary systems comprising of a FGK-type primary and an M dwarf secondary provide a powerful solution to this calibration problem. The metallicity of the FGK star, which can be accurately derived using well-established spectroscopic methods \citep[e.g.][]{Lee-2008AJ....136.2022L,Lee-2008AJ....136.2050L,Smolinski-2011AJ....141...89S,Du-2019ApJS..240...10D} 
can be reliably assumed to be the same as that of the M dwarf companion due to their common origin. This co-evolution makes such systems ideal benchmarks for calibrating the metallicity of M dwarfs \citep[e.g.][]{Rojas-Ayala-2010,Mann-2013a,Montes-2018,Souto-2020ApJ...890..133S,Souto-2022ApJ...927..123S,Qiu-2024MNRAS.52711866Q}. 

In this work, we analyzed over 1100 FGK+M wide binaries. The FGK dwarf primaries are from LAMOST DR11 \citep{Cui-2012,Deng-2012,Zhao-2012} and M dwarf secondaries are from the SDSS-V/BOSS. We developed the SLAM model as one of the pipelines that SDSS-V uses to calibrate the stellar parameters for all BOSS M dwarfs. This pipeline can be used to determine the \feh, \teff, \logg\ and \alp\, of BOSS M dwarfs. The SLAM is a data-driven method based on Support Vector Regression (SVR). As a robust nonlinear regression technique, SVR has been extensively utilized in various fields of astronomy \citep{Liu-2012,Liu-2015}, particularly in spectral data analysis \citep{Li-2014,Liu-2014,Bu-2015}. The SLAM has demonstrated excellent performance in determining stellar atmospheric parameters from spectra \citep{Zhang-2020,Li-2021,Qiu-2024MNRAS.52711866Q}. 

This paper is organized as follows: Section \ref{sect:Data} describes the selection of M dwarf candidates and the identification of FGK+M and M+M binary systems. Section \ref{sect:method} details the training and prediction process of the SLAM model. In Section \ref{sect:result}, we analyze the SLAM-predicted results and compare them with those from other literature. We also calibrate the ASPCAP's \feh\ of M
dwarfs in Section \ref{sect:cali_ASPCAP}. Finally, Section \ref{sect:conclu} summarizes the methodology and findings of this study.

\section{Data} \label{sect:Data}
We describe SDSS-V/BOSS data briefly in Subsection \ref{sect:boss-data}. The method of identification of BOSS M dwarfs, as shown in Subsection \ref{sect:Identification of M dwarfs}, adopts a combination of the BOSS spectral pipeline and Gaia photometric data. The selection of FGK+M and M+M wide binary systems was detailed in Subsection \ref{sect:binary}, where we employed astrometric and photometric criteria to ensure robust binary identification.

\subsection{SDSS-V/BOSS data}\label{sect:boss-data}

The SDSS/BOSS spectrographs, initially designed for the Baryon Oscillation Spectroscopic Survey (BOSS) in SDSS-III\citep{Dawson-2013}, have been upgraded and repurposed for SDSS-V while maintaining their fundamental capabilities. 
In SDSS-V, the twin BOSS spectrographs, which are located at APO and LCO, respectively, support an all-sky, multi-object fiber spectroscopic capability, enabling large-scale stellar, Galactic, and extragalactic studies.  

The 1d analysis portion of the BOSS spectroscopic reduction pipeline \citep[idlspec2d, v6\_1\_3;][Morrsion et al.\textit{in prep}]{Bolton-2012AJ....144..144B} uses chi-squared minimization template fitting to first determine quasar, galaxy, and star classification. The stellar classification is further subclassed using 117 stellar archetype templates built from the Indo-US spectral database \citep{Valdes-2004ApJS..152..251V}, models from the POLLUX spectrum database \citep{Palacios-2010A&A...516A..13P}, 6 stellar archetype PCA templates, and a set of 3 cataclysmic variable star PCA eigenspectra built from early BOSS data. After fitting these models, the best fit is reported as the class and subclass of the targets.

\subsection{M dwarf candidates}\label{sect:Identification of M dwarfs}

\begin{figure}
\centering
\includegraphics[width=0.5\textwidth, trim=0.cm 0.0cm 0.0cm 0cm, clip]{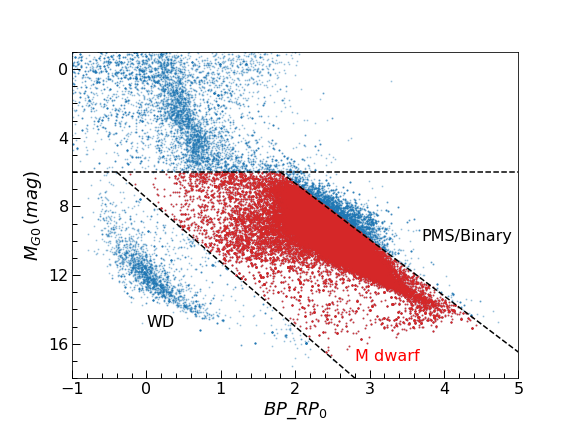}
\caption{The color-absolute magnitude diagram ($BP\_RP_{0}$ vs. $M_{G0}$) of all stars classified as M-type stars by the BOSS spectroscopic reduction pipeline in DR19. The blue points represent objects that do not satisfy the criteria 1–4 defined in Section \ref{sect:Identification of M dwarfs}, and are therefore considered non-M dwarfs. Red points denote M dwarf candidates identified according to criteria 1–4.} \label{fig:CMD_M}
\end{figure}

 We identified the M dwarf candidates based on the targets which were labeled as M-type stars by the BOSS spectroscopic reduction pipeline in data release 19 (DR19). The reddening value $E(B-V)$ for each star is obtained from the three-dimensional dust map \citep{Green-2019}. The extinction value of $A_{\rm V}$ can be calculated by $A_{\rm V}=3.1\times E(B-V)$ \citep{Cardelli-1989ApJ...345..245C}. The extinction values of Gaia's three bands can be transferred from $A_{\rm V}$ to $A_{\rm G}, A_{\rm BP}, A_{\rm RP}$ by using the extinction coefficient provided by \citet{Gaia-2018a}. Therefore, the color-absolute magnitude diagram (CMD) of these objects which were labeled as M-type stars by BOSS spectroscopic reduction pipeline is shown in Figure \ref{fig:CMD_M}. $M_{\rm G0}$ is the absolute magnitude of Gaia $G$ band, $M_{\rm G0}=phot\_g\_mean\_mag+5*\log_{10}(\varpi/{\rm mas})-10-A_{\rm G}$. $ BP\_RP_{0}$ is the color index of Gaia $BP$ and $RP$ bands, $BP\_RP_{0}$=$(phot\_bp\_mean\_mag-A_{\rm BP})-(phot\_rp\_mean\_mag-A_{\rm RP}$). The $phot\_g\_mean\_mag$, $phot\_bp\_mean\_mag$ and $phot\_rp\_mean\_mag$ are the apparent magnitudes of Gaia $G, BP$ and $RP$ bands, respectively. $\varpi$ is the Gaia parallax.
 Apparently, the diagram reveals the presence of non-M dwarfs within the sample. To remove these contaminants, we applied empirical criteria as follows:

\begin{enumerate}
    \item \texttt M$\rm_{G0}$ \textgreater 6.
    \item \texttt M$\rm_{G0}$ \textless 3.75*($\rm BP\_RP_{0}$)+7.50.
    \item \texttt M$\rm_{G0}$ \textgreater 3.28*($\rm BP\_RP_{0}$)+0.10.
    \item \texttt ruwe \textless 1.4.
    
\end{enumerate}
The first criterion is designed to remove bright stars, such as F, G, or K dwarfs, which may contaminate the M dwarf sample. The second criterion, as empirically defined by \citep{Birky-2020,Li-2021,Qiu-2024MNRAS.52711866Q}, is set to exclude pre-main sequence stars and unresolved binary systems, while the third criterion aims to remove white dwarfs. The fourth criterion uses the re-normalized unit weight error (ruwe) from Gaia to further exclude unresolved binaries \citep{Penoyre-2022MNRAS.513.5270P} or sources with poor and/or variable astrometric data. Finally, there are over $\sim$90,000 M dwarf candidates meet above four criteria, as displayed by the red dots in Figure \ref{fig:CMD_M}.

\subsection{FGK+M and M+M wide binary}\label{sect:binary}
We started with the wide binary catalog of \citet{El-Badry-2021}, which identified wide binaries within 1 Kpc in Gaia eDR3 \citep{Gaia-2021}. The potential binary candidates were identified by searching for pairs of stars with projected separation less than 1 parsec. Then they imposed strict cuts on proper motion and parallax differences between the components to ensure the two stars in a pair are physically bound. After eliminating clusters, background pairs, and triples, their final catalog provided a robust and uncontaminated sample with 1,817,594 wide binaries.

The Large Sky Area Multi-Object Fiber Spectroscopic Telescope \citep[LAMOST;][]{Cui-2012,Deng-2012,Zhao-2012} has delivered approximately 7.4 million spectra of F-, G-, and K-type dwarfs in its A/F/G/K stellar catalog in Data Release 11 (DR11). Based on LAMOST DR5 observations, \citet{Du-2019ApJS..240...10D} constructed an empirical stellar spectral library for F/G/K and late-type A stars, achieving an internal precision of 0.05 dex and an external accuracy of 0.1 dex in [Fe/H], thereby providing a robust basis for reliable stellar parameter determination.


We cross-matched the LAMOST F, G and K dwarfs with the final wide binary catalog of \citet{El-Badry-2021} and identified the M dwarf secondaries using the criteria 1-4 outlined in Subsection \ref{sect:Identification of M dwarfs}. This process yielded 1,665 FGK+M wide binaries. To ensure data quality, we applied additional criteria to remove binaries with poor-quality data:  \\
1) Metallicity Uncertainty: $\rm [Fe/H]_{FGK,err}<0.2$,  $\rm [Fe/H]_{FGK,err}$ is the metallicity uncertainty of the LAMOST F/G/K dwarfs. This criterion ensures that the \feh\ values of the primary stars are of reasonable quality.\\
2) Chance Alignment Probability: R\_chance\_align \textless 0.1, where R\_chance\_align\footnote{It is evaluated in a seven-dimensional space as described in \citep{El-Badry-2021}.} represents the probability that a wide binary is a chance alignment. High-confidence binaries are expected to have low R\_chance\_align values, with R\_chance\_align$<$0.1 corresponding to a \textgreater 90\% probability of being physically bound.\\
3) Data Quality: $\rm ruwe_1<1.4$ and $\rm ruwe_2<1.4$. The $\rm ruwe_1$ and $\rm ruwe_2$ are the re-normalised unit weight error of FGK dwarf primaries and M dwarf secondaries, respectively. This criteria is consistent with the fourth criterion described in Subsection \ref{sect:Identification of M dwarfs}. It was set to exclude unresolved binaries as well as sources with poor and/or variable astrometric data.\\
Applying these selections results in a total of 1120 FGK+M wide binaries remaining. A comprehensive list of these binaries can be found in Table \ref{table:binary}.

\begin{figure}
\centering
\includegraphics[width=0.5\textwidth, trim=0.cm 0.0cm 0.0cm 0cm, clip]{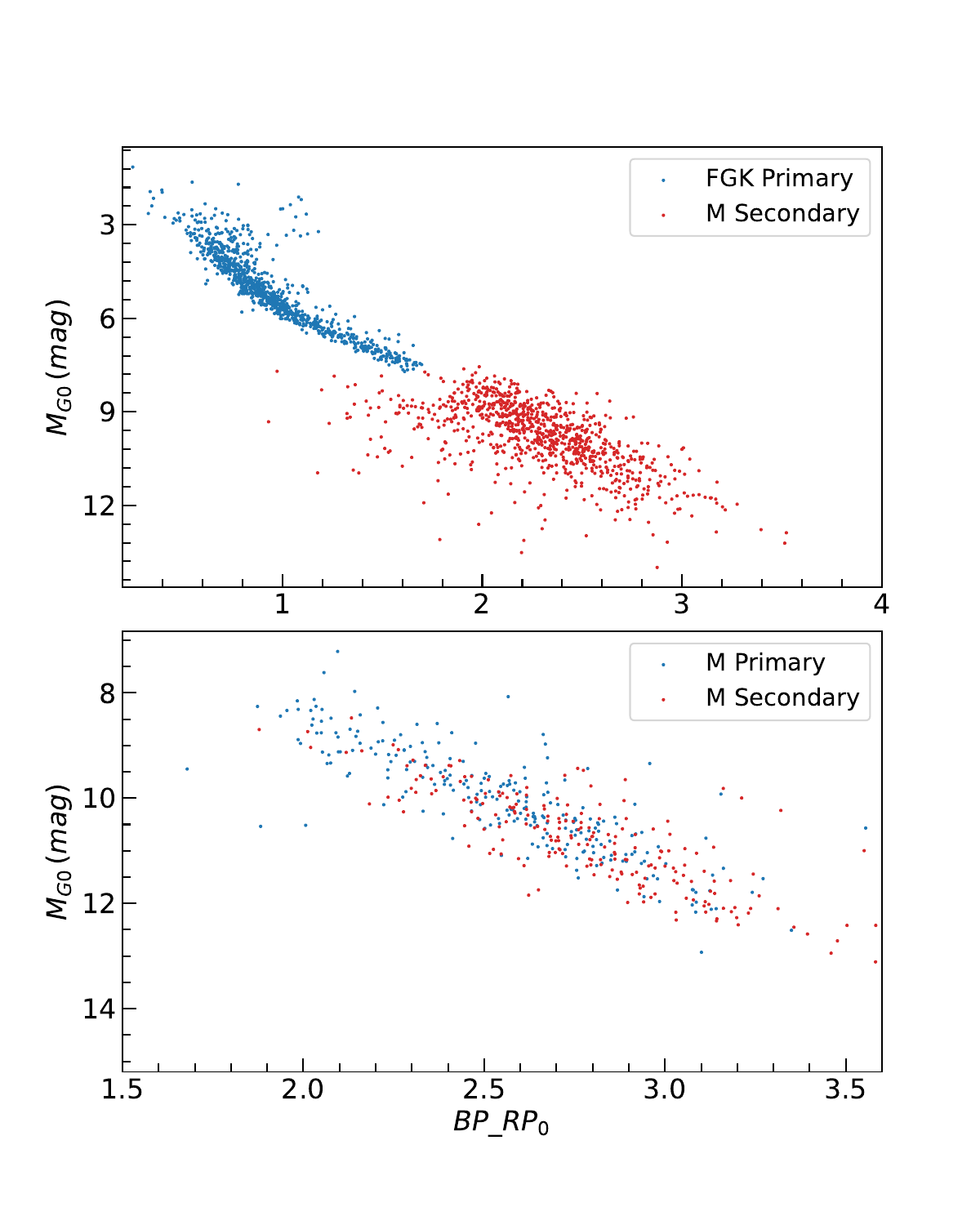}
\caption{The top panel is the CMD of 1120 FGK+M wide binaries. The blue and red dots are the LAMOST FGK primaries and BOSS M secondaries, respectively. The bottom panel is the same as the top panel, but for the M+M wide binaries. The blue and red dots represent the M dwarf primaries and secondaries, respectively.}\label{fig:params}
\end{figure}

Furthermore, we identified 256 BOSS M+M wide binaries with R\_chance\_align $<$ 0.1 and ruwe $<$1.4 for both M components. The M+M wide binary sample serves as a validation set for the SLAM predicted parameters, as described in Subsection \ref{sec:M+M_validate}. Figure \ref{fig:params} shows the CMDs of two components of FGK+M (top panel) and M+M (bottom panel) wide binaries.

\subsection{Training and test dataset}\label{sect:training data}

We adopted the \feh\ of 1120 M dwarfs from the LAMOST DR11 F, G or K dwarf companions. 
For the \teff\ and \logg\ of 1120 M dwarfs, we considered the values derived from APOGEE Net (III) \citep{Sizemore-2024AJ....167..173S}, which is a data–driven neural–network pipeline that infers stellar parameters from APOGEE near–infrared spectra using labels assembled from a variety of external sources. The first two generations, APOGEE Net I \citep{Olney-2020AJ....159..182O} and APOGEE Net II \citep{Sprague-2022AJ....163..152S}, established the basic framework by training on The Payne catalog \citep{Ting-2019ApJ...879...69T} of giants and dwarfs, supplemented with several thousand M dwarfs and young stellar objects (YSOs); these models deliver typical uncertainties of $\sim150$–$180$ K in $T_{\rm eff}$, $\sim0.2$ dex in $\log g$ for APOGEE Net I, and $\sim30$ K, $\sim0.03$ dex, respectively, for APOGEE Net II. Building on this foundation, APOGEE Net III \citep{Sizemore-2024AJ....167..173S} reassembles a much larger and more heterogeneous training set, combining labels from APOGEE Net II, BOSS Net \citep{Sizemore-2024AJ....167..173S}, The Payne \citep{Ting-2019ApJ...879...69T}, The zeta Payne
catalog \citep{Straumit-2022AJ....163..236S}, Gaia \citep{Fouesneau-2023A&A...674A..28F}, and several specialized catalogs \citep[e.g,][]{McBride-2021AJ....162..282M} to yield $\sim4\times10^{5}$ labeled stars spanning very cool stars ($T_{\rm eff}<3000$ K) to hot stars ($T_{\rm eff}\, \sim 5\times 10^{4}$ K). Quantitatively, the uncertainties in $T_{\rm eff}$, $\log g$ for APOGEE Net III is very similar to that of APOGEE Net II. This updated model is applied to all legacy SDSS-I–IV and SDSS-V APOGEE spectra, providing homogeneous, near-IR stellar parameters with performance broadly consistent with APOGEE Net II while significantly extending the reliable coverage to both the coolest and the hottest parts of the parameter space.

\begin{figure*}
\centering
\includegraphics[width=1\textwidth, trim=0.cm 0.0cm 0.0cm 0cm, clip]{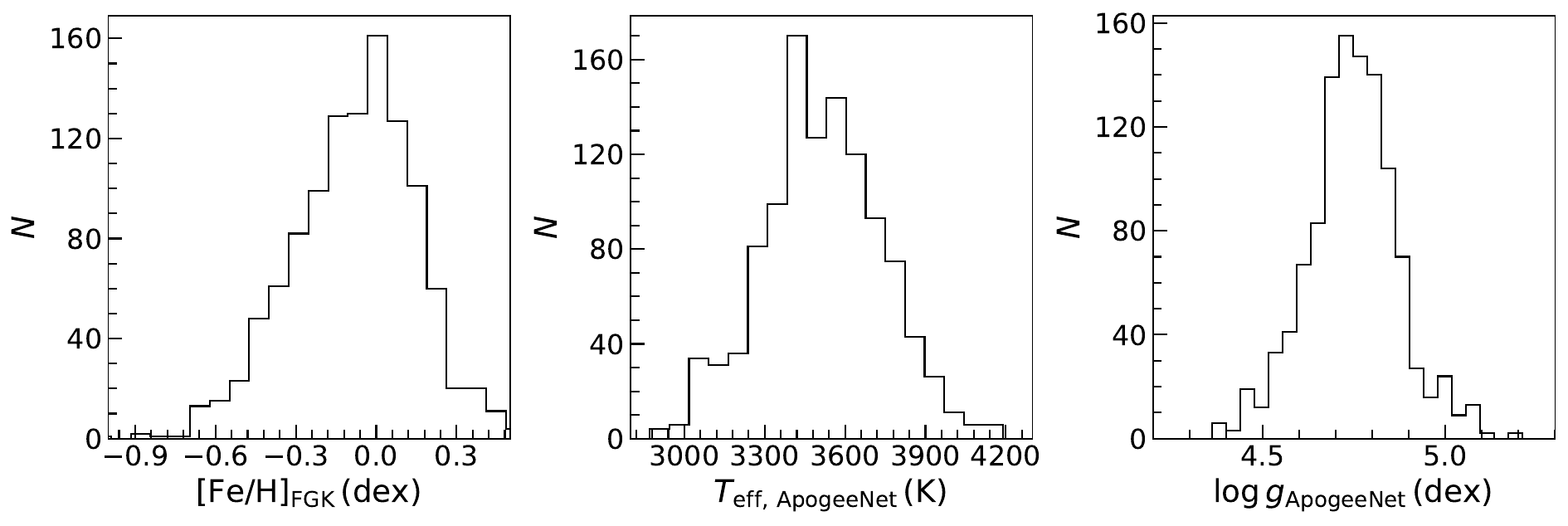}
\caption{The distributions of SLAM training labels, including \feh, \teff\ and \logg, which described in Section \ref{sect:training data}. The ranges of \feh,  \teff\ and \logg\ are (-1, 0.5) dex, (2900,4200) K and (4.4,5.1) dex, respectively.
}\label{fig:train_params}
\end{figure*}

 We randomly divided the 1,120 M dwarfs in wide binary systems into a training set of 870 stars and a test set of 250 stars. The training dataset for the SLAM model consists of 870 M dwarf spectra from SDSS-V/BOSS, along with their corresponding parameters, namely \feh, \teff, and \logg. The test set includes 250 BOSS M dwarf spectra and the corresponding parameters. Figure \ref{fig:train_params} displays the distribution of the three training labels. The parameters cover the following ranges: \feh\, from -1 to 0.5 dex, \teff\, from 2900 to 4200 K, and \logg\, from 4.4 to 5.1 dex. It is worth noting that the number of training stars outside the ranges \feh\ $\in[-0.6,0.5]$ dex, \teff\ $\in [3100,3900]$ K, and \logg\ $\in[4.45,4.95]$ dex is small ($<$ 20 objects), so the uncertainties of SLAM predictions in these sparsely sampled regions may exceed the nominal values.
 

\begin{table}
\centering
\caption{Parameter coverage of the SLAM training set.}\label{table:train_range}
\begin{tabular}{lll}
\hline 
Parameter & Total range  & Usable range \\ 
 & (training set) & (SLAM valid) \\ 
\hline
[Fe/H] (dex) & [-1, 0.5] & [-0.6, 0.5]   \\ 
\teff (K) & [2900, 4200] & [3100, 3900] \\
\logg (dex) & [4.4, 5.1] & [4.45, 4.95]\\
\hline 
\end{tabular}

\medskip
\begin{flushleft}\footnotesize
\textit{Note.} ``Total range'' denotes the full span of parameters in the assembled training sample, while ``Usable range'' indicates the range over which SLAM is validated and recommended for scientific use.
\end{flushleft}
\end{table}

\section{Method}\label{sect:method}
As in our previous works \citep{Qiu-2023RAA....23e5008Q,Qiu-2024MNRAS.52711866Q}, we adopted the SLAM, which is a data-driven method based on Support Vector Regression (SVR), to determine the parameters of M dwarfs. The training and prediction processes of SLAM are detailed in Subsection \ref{sect:Train_data} and \ref{sect:Pre_data}, respectively. The determination of parameter uncertainties as described in 
Subsection \ref{sect:error}.

\subsection{Model Training} \label{sect:Train_data}   

Before training the SLAM model, we normalized the BOSS M dwarf spectra. A smoothing spline \citep{de-1977SJNA...14..441D} was applied to smooth the entire spectrum and derive the pseudo-continuum. The observed spectra were then normalized by dividing them by their corresponding pseudo-continua. The normalized spectra and stellar labels also need to be standardized, so that both stellar labels and spectral fluxes have a mean of 0 and a variance of 1. The details about the data preprocessing can be found in subsection 2.1 in \citet{Zhang-2020}.

Let $\boldsymbol{\vec{\theta}}_i$ represent the stellar label vector of the $i$th star in the training set. The $j$th pixel of the training spectrum and the model output spectrum corresponding to the star with stellar label vector $\boldsymbol{\vec{\theta}}_i$ is denoted as \emph{f$_j$}($\boldsymbol{\vec{\theta}}i$) and $f{i,j}$, respectively. The mean squared error (MSE) and median deviation (MD) for the $j$th pixel can be calculated using a specific set of hyperparameters. In SLAM, the radial basis function (RBF) is the kernel for SVR. The hyperparameters $C$, $\varepsilon$, and $\gamma$, which correspond to the penalty level, tube radius, and width of the RBF kernel, respectively, are automatically optimized for each pixel based on the training set. The MSE and MD are shown in equations (\ref{eq:MSE}) and (\ref{eq:MD}).

\begin{equation}\label{eq:MSE}
MSE_\emph{j}=\frac{1}{m}\sum_{i=1}^m\big[\emph{f$_j$}\big(\boldsymbol{\vec{\theta}}_i\big)-\emph{f$_ {i,j}$}\big]^2.
\end{equation}

\begin{equation}\label{eq:MD}
MD_\emph{j}=\frac{1}{m}\sum_{i=1}^m\big[\emph{f$_j$}\big(\boldsymbol{\vec{\theta}}_i\big)-\emph{f$_{i,j}$}\big].
\end{equation}
\\
Theoretically, smaller MSE and MD values indicate better fitting. However, if the SLAM model is trained using the entire training set, it may result in an overfitted model, where $\rm MSE_\emph{j}$ and $\rm MD_\emph{j}$ are both reduced to zero. To prevent overfitting, \citet{Zhang-2020} employed k-fold cross-validated MSE (CV MSE) and k-fold cross-validated MD (CV MD) to evaluate $\rm MSE_\emph{j}$ and $\rm MD_\emph{j}$. In this approach, the training set is randomly divided into k subsets, with k$=$10 in this study. The predicted value \emph{f$_j$}($\boldsymbol{\vec{\theta}}i$) is generated using a model trained on the remaining k-1 subsets. The optimal hyperparameters for SVR at each pixel are determined by identifying the hyperparameter set that minimizes the CV MSE$_{j}$ during the search process.

\subsection{Model Prediction} \label{sect:Pre_data} 
The posterior probability density function of the stellar label vector for a given observed spectrum is expressed using the Bayesian formula, as shown in equation (\ref{eq:PDF}),

\begin{equation}\label{eq:PDF}
p\big(\boldsymbol{\vec{\theta}}\mid\boldsymbol{\vec{f}}_{obs}\big)\propto p\big(\boldsymbol{\vec{\theta}}\big)\prod_{j=1}^n p\big(f_{j,obs}\mid\boldsymbol{\vec{\theta}}\big).
\end{equation}
where $\boldsymbol{\vec{\theta}}$ denotes the stellar label vector, while $\boldsymbol{\vec{f}}_{obs}$ and \emph{$f_{j,obs}$} represent the normalized observed spectrum vector and the normalized flux at the \emph{j}th pixel of the observed spectrum, respectively. The term $p(\boldsymbol{\vec{\theta}})$ refers to the prior probability of the stellar label vector $\boldsymbol{\vec{\theta}}$, and $p(f_{j,obs}\mid\boldsymbol{\vec{\theta}})$ is the likelihood of the observed spectrum flux at the \emph{j}th pixel (\emph{$f_{j,obs}$}) given the stellar label vector $\boldsymbol{\vec{\theta}}$. The stellar labels can be determined by maximizing the posterior probability $p(\boldsymbol{\vec{\theta}}\mid\boldsymbol{\vec{f}}_{obs})$.

\subsection{Parameter uncertainties}\label{sect:error}
Equations (\ref{eq:bias}) and \ref{eq:scatter}) define the cross-validated bias ($\rm CV\_bias$) and cross-validated scatter ($\rm CV\_scatter$) of stellar labels, analogous to the CV MD and CV MSE of the spectrum discussed in 
Subsection \ref{sect:Train_data}. That is, once the SLAM model predicts the stellar labels for the stars with known ground truth stellar parameters, the $\rm CV\_bias$ and $\rm CV\_scatter$ of these stars can be calculated. These metrics can be considered as the mean deviation and standard deviation of the stellar labels, respectively, and are used to evaluate the precision of the stellar parameters derived from the SLAM model. In general, smaller $\rm CV\_scatter$ and $\rm CV\_bias$ values indicate a better-trained model.

\begin{equation}\label{eq:bias}
CV\_bias=\frac{1}{m}\sum_{i=1}^m\left(\boldsymbol{\vec{\theta}}_{i,SLAM}-\boldsymbol{\vec{\theta}}_{i}\right).
\end{equation}

\begin{equation}\label{eq:scatter}
CV\_scatter=\frac{1}{m}\sqrt{\sum_{i=1}^m\left(\boldsymbol{\vec{\theta}}_{i,SLAM}-\boldsymbol{\vec{\theta}}_{i}\right)^2}.
\end{equation}

\section{Experiments and Results}\label{sect:result}
 The performance of the trained SLAM model on the test dataset is shown in Subsection \ref{sec:apply_SLAM}. We also explore the uncertainties of SLAM predicted parameters as a function of spectral signal-to-noise ratio ($\mathrm {SNR}$) in Subsection \ref{sec:err_eq}. Some comparisons in stellar parameters between this work and the literature are exhibited in Subsection \ref{sect:Validation}. 

\subsection{SLAM model training and performance} \label{sec:apply_SLAM}
\begin{figure*}
\centering
\includegraphics[width=1\textwidth, trim=0.cm 0.0cm 0.0cm 0cm, clip]{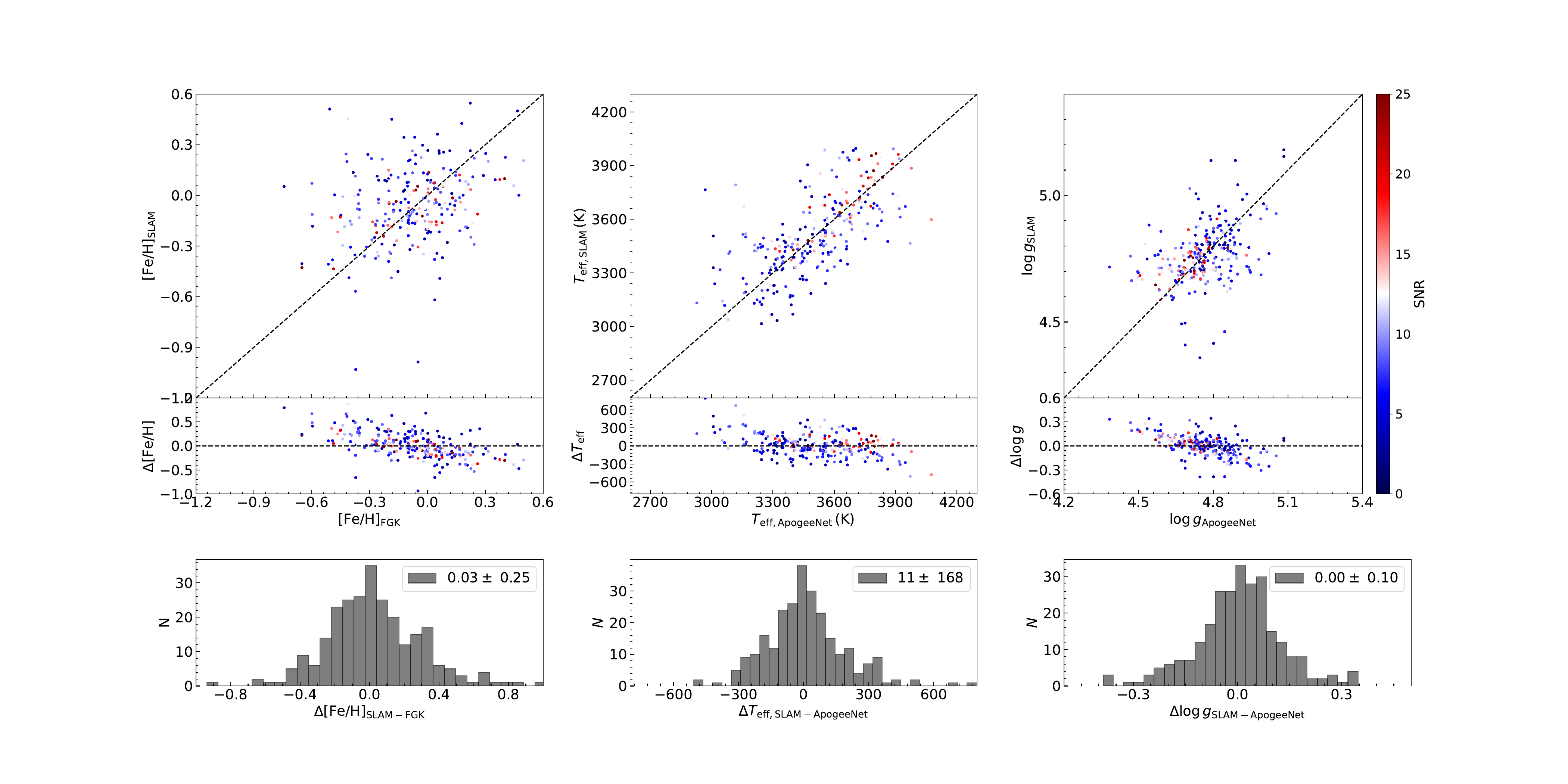}
\caption{ Top row: one-to-one comparisons between SLAM prediction values and the reference labels for \feh\ (FGK catalogue; left), \teff\ (APOGEE Net; middle), and \logg\ (APOGEE Net; right); points are colour–coded by BOSS spectral $\mathrm {SNR}$. Middle row: corresponding residuals, $\Delta[{\rm Fe/H}]_{\rm SLAM-FGK}$, $\Delta T_{\rm eff,SLAM-ApogeeNet}$, and $\Delta\log g_{\rm SLAM-ApogeeNet}$, as a function of the reference labels, with dashed lines marking zero offset. Bottom row: distributions of these residuals, with text boxes giving the median bias and standard deviation of difference (approximately $0.03\pm0.25$ dex in \feh, $11\pm168$ K in \teff, and $0.00\pm0.10$ dex in \logg).}\label{fig:test_data}
\end{figure*}

We trained the SLAM model with 870 normalized M dwarf spectra and the corresponding stellar labels. 
Figure \ref{fig:test_data} exhibits the comparisons of three parameters between SLAM predictions (y-axis) and the reference values (x-axis) of 250 test stars. From left to right, panels correspond to the comparisons of \feh, \teff, and \logg. There are no obvious trends with the benchmark values across the sampled ranges, and the residual distributions are approximately symmetric with small biases. Quantitatively, the median offsets and standard deviation are $\sim0.03\pm0.25$ dex in \feh, $11\pm168$ K in \teff, and $0.00\pm0.10$ dex in \logg, demonstrating that SLAM reproduces the reference scale at the level required for our analysis. It indicates that the predicted parameters derived from the trained SLAM model are reliable. Furthermore, a clear $\mathrm {SNR}$ dependence is visible, low–SNR spectra (blue points) exhibit noticeably larger scatter in all three parameters, while high–SNR spectra (red points) concentrate more tightly around the one–to–one relation and $\Delta=0$, indicating that SLAM precision improves with data quality.

\subsection{Determination of the parameter uncertainties} \label{sec:err_eq}

\begin{figure*}
\centering
\includegraphics[width=1\textwidth, trim=0.cm 0.0cm 0.0cm 0cm, clip]{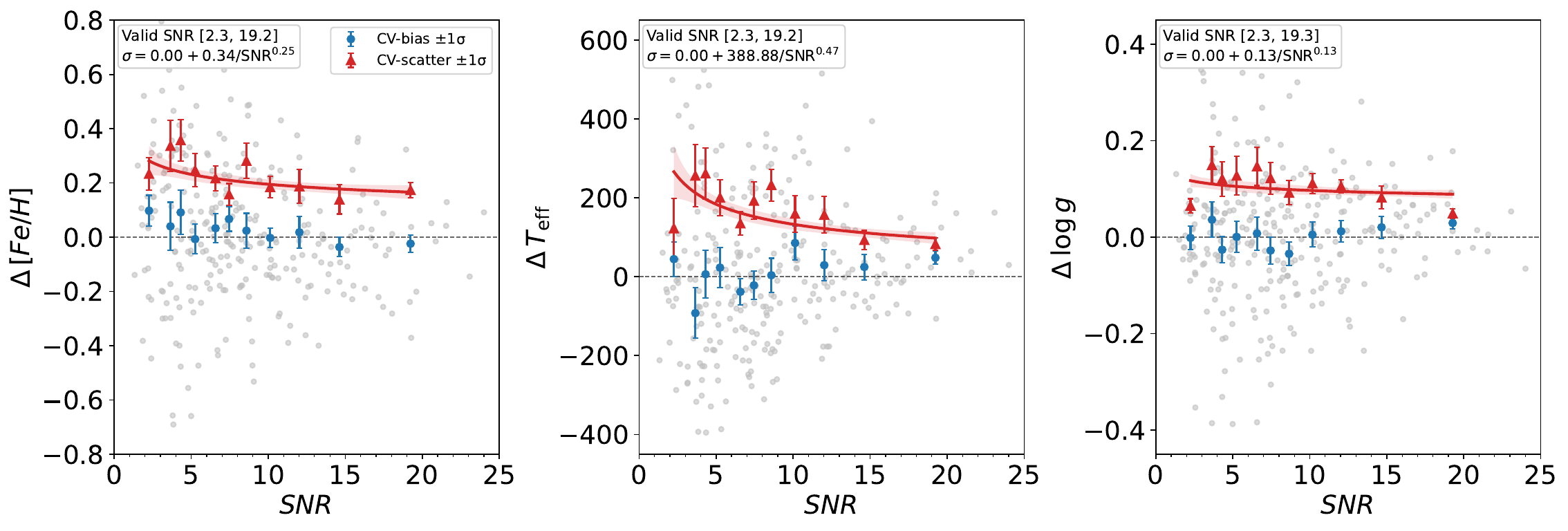}
\caption{The left panel shows the $\rm \Delta [Fe/H] (=[Fe/H]_{SLAM-FGK})$ versus the spectral $\mathrm {SNR}$ of the test dataset. 
Gray points are individual cross‐validation residuals for single stars. Stars are sorted by $\mathrm {SNR}$ and grouped into equal‐count $\mathrm {SNR}$ bins ($\sim$20); in each bin the blue circles with error bars show the median $\mathrm {CV\_bias}$ and its 1$\sigma$ uncertainty, while the red triangles with error bars show the median $\mathrm {CV\_scatter}$ and its uncertainty. The solid red curves are weighted fits to the binned scatters of the form $\sigma=a+b/\mathrm {SNR}^{c}$, with the best‐fitting relation and the valid $\mathrm {SNR}$ range indicated in the upper part of each panel. The middle and right panels are the same as the left panel, but for \teff\  and \logg\ , respectively.}\label{fig:err}
\end{figure*}

\citet{Zhang-2020} reported that the formal uncertainties returned by the SLAM model are underestimated compared with the $\rm CV\_scatter$ values. Therefore, in this work, we adopt $\mathrm{CV\_scatter}$ as our empirical estimate of the parameter uncertainties, as defined in Section~\ref{sect:error}. 
Figure \ref{fig:err} displays the $\mathrm {CV\_bias}$ (blue symbols) and $\mathrm {CV\_scatter}$ (red symbols) versus the spectral $\mathrm{SNR}$, using equal-count SNR bins of the test data set (at least $\sim 20$ stars per bin), where the $\mathrm{SNR}$ values are taken from the BOSS pipeline. The $\mathrm {CV\_bias}$ are consistent with zero over the full SNR range, while the $\mathrm {CV\_scatter}$ decrease as the spectral SNR increases. We fit the CV scatters as a functions of SNR for the three labels and adopt the following relations:
\begin{align}
\sigma_{[\mathrm{Fe/H}]} &= 0.34\times \,\mathrm{SNR}^{-0.25}, \\
\sigma_{T_{\mathrm{eff}}} &= 388.88\times \,\mathrm{SNR}^{-0.47}\ \mathrm, \\
\sigma_{\log g} &= 0.13\times \,\mathrm{SNR}^{-0.13}.
\end{align}
These fits are calibrated over the SNR range covered by the equal-count bins, i.e. $\mathrm{SNR}\simeq 2.3$–$19.2$ for [Fe/H] and $T_{\mathrm{eff}}$, and $\mathrm{SNR}\simeq 2.3$–$19.3$ for $\log g$; for spectra with SNR outside these intervals the above formulae represent extrapolations and the true uncertainties may differ from the nominal values.
At $\mathrm{SNR}=15$, these relations correspond to typical uncertainties of
$\sim 0.19$\,dex in [Fe/H], $\sim 132$\,K in $T_{\mathrm{eff}}$, and
$\sim 0.1$\,dex in $\log g$.

\subsection{SLAM predicted parameters validation}\label{sect:Validation}
We applied the trained SLAM model to all the M dwarfs discussed in Subsection \ref{sect:Identification of M dwarfs} to derive their stellar parameters. The uncertainties of these parameters refer to Subsection \ref{sec:err_eq}. We list some comparison results of three parameters between this work and other literature in this section. The SLAM parameter distributions of all the M dwarfs are shown in Setion \ref{sect:Validation} and \ref{sect:cali_ASPCAP}.

\subsubsection{\feh\ comparison}\label{sec:M+M_validate}
\begin{figure*}
\centering
\includegraphics[width=1\textwidth, trim=0.cm 0.0cm 0.0cm 0cm, clip]{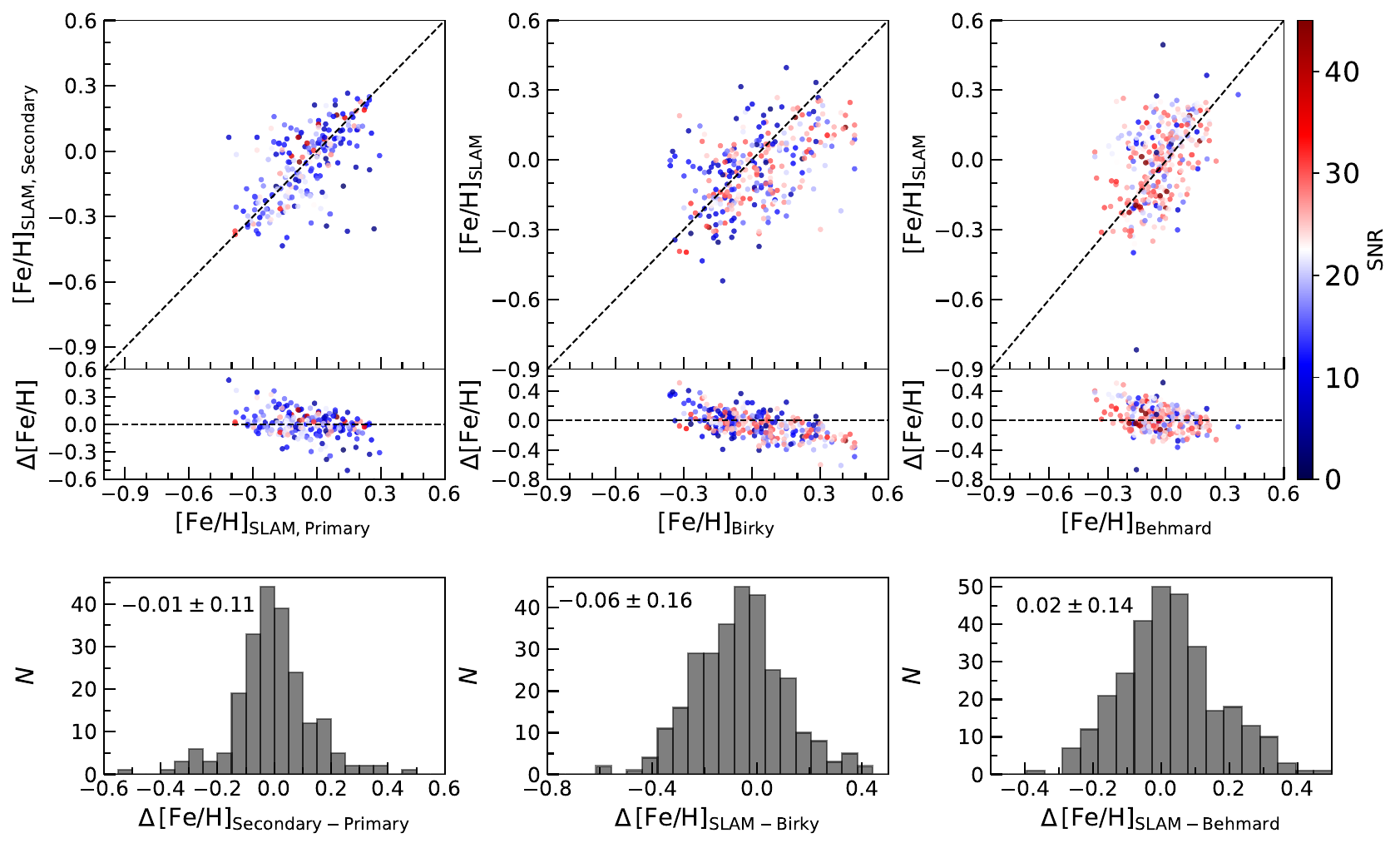}
\caption{
Left panels: The top-left panel compares the SLAM-predicted metallicities for primary M dwarfs ($\rm[Fe/H]_{Primary}$) and secondary M dwarfs ($\rm[Fe/H]_{Secondary}$) in M+M wide binaries. The points are colored by the spectral $\mathrm {SNR}$, as indicated by the color bar on the right. A residual plot is shown below the main comparison. The bottom-left panel displays the distribution of the metallicity difference, $\rm \Delta [Fe/H]$=$\rm[Fe/H]_{Secondary}-[Fe/H]_{Primary}$. Middle panels and right panels are the same as the left panels, but compare the SLAM-predicted metallicities with those from \citet{Birky-2020} ($\rm [Fe/H]_{Birky}$) and \citet{Behmard-2025arXiv250114955B} ($\rm [Fe/H]_{Behmard}$), respectively.
}\label{fig:SLAM_vali}
\end{figure*}

We utilized M+M wide binary systems to validate the \feh\ derived from the SLAM model, under the assumption that both components of a physically bound binary should exhibit identical metallicities. The left panels in Figure \ref{fig:SLAM_vali} show the comparison in metallicity between two components for the M+M wide binaries described in Subsection \ref{sect:binary}. The \feh\ of the two components ($\rm [Fe/H]_{slam,primary}$ and $\rm [Fe/H]_{slam,secondary}$) are derived from the SLAM model independently. The results indicate there is no significant bias (-0.01 $\pm$ 0.11 dex), demonstrating the consistency of the SLAM-derived metallicities for physically bound pairs.

We further compared the metallicities from our work ($\rm [Fe/H]_{slam}$) with those of \citet{Birky-2020} ($\rm [Fe/H]_{Birky}$) and \citet{Behmard-2025arXiv250114955B}($\rm [Fe/H]_{Behmard}$), as shown in the middle and right panels of Figure \ref{fig:SLAM_vali}. Both studies utilized the data-driven method, $The \,\,Cannon$\citep{Ness-2015ApJ...808...16N}, to determine \feh\ of M dwarfs based on high-resolution near-infrared spectra from the APOGEE survey. Their analyses relied on samples of M dwarfs in wide binaries with FGK-type primaries as metallicity benchmarks, ensuring that the metallicity estimates in both works were calibrated using F, G, and K dwarf companions. The median and standard deviation values of the differences $\rm \Delta [Fe/H]_{slam-Birky}$ and $\rm \Delta [Fe/H]_{slam-Behmard}$ are -0.06 $\pm$ 0.16 dex and 0.02 $\pm$ 0.14 dex, respectively. In these three panels, the significant discrepancies observed in some sources (blue points) are caused by low $\mathrm {SNR}$ of their BOSS spectra. All of these comparisons demonstrate the reliability of the \feh\ values derived from the SLAM model.

\subsubsection{\teff\ comparison}

\begin{figure*}
\centering
\includegraphics[width=1.\textwidth, trim=0.cm 0.0cm 0.0cm 0cm, clip]{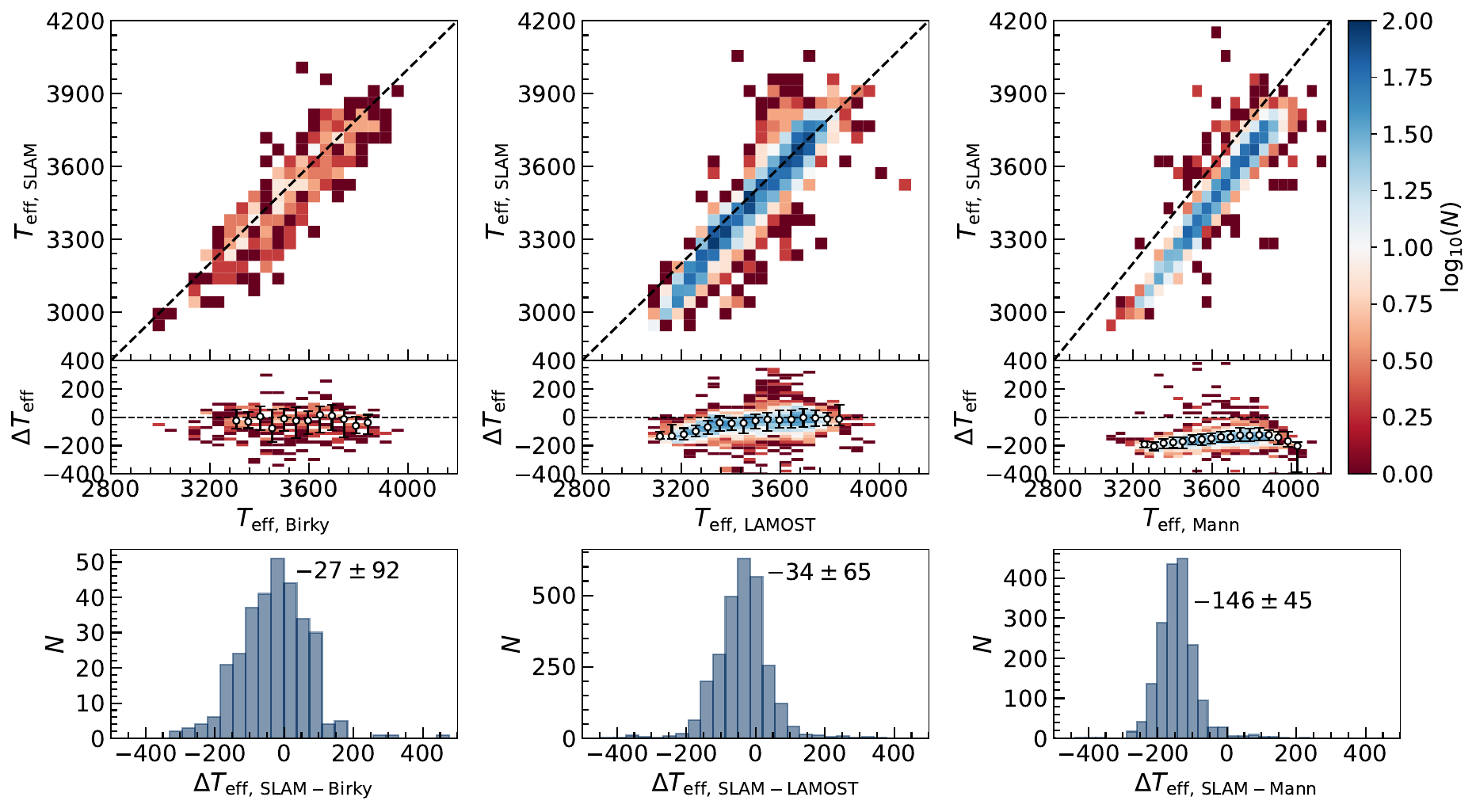}
\caption{ A similar plot as Figure \ref{fig:SLAM_vali}. This figure presents a comparison of the effective temperatures predicted by SLAM ($\rm T_{eff,slam}$) with those reference \teff, i.e., $\rm T_{eff,Birky}$ (\citet{Birky-2020}, left), $\rm T_{eff,LAMOST}$, (LAMOST, middle) and $\rm T_{eff,Mann}$, (\citet{Mann-2015,Mann-2016}, right), respectively. Three top panels show 2D density plot of $T_{\rm eff,slam}$ versus $ T_{\rm eff,ref}$. Three middle panels present the temperature difference as a function of $T_{\rm eff,ref}$. The white circles and black color bars mark the median and standard deviation of residuals in different temperature bins. The histograms in the three bottom panels show the distribution of the temperature difference ($\Delta T_{\rm eff}=T_{\rm eff,slam}$-$ T_{\rm eff,ref}$), with a median and standard deviation of -27 $\pm$ 92 K, -34$\pm$65, and -146$\pm$45 K, respectively. 
  }\label{fig:SLAM_teff}
\end{figure*}

For the effective temperature (\teff), we compared the SLAM predictions ($ T_{\rm eff,slam}$) with those of \citet{Birky-2020} ($ T_{\rm eff,Birky}$), who firstly derived the effective temperatures of M dwarfs by comparing optical spectra to BT-Settl atmospheric models and then calibrated using interferometric angular diameters of 29 M dwarfs. The comparisons are shown in the left panels of Figure \ref{fig:SLAM_teff}. There is a systematic difference of -27$\pm$92 K. Additionally, we compared  $ T_{\rm eff,slam}$ with the \teff\ values from the LAMOST gM/dM/sdM stars catalog (\citet{Du-2024ApJS..275...42D},middle panels), they built a clean, homogeneous M-dwarf label set to reduce systematics in spectroscopic parameters. After Gaia-based calibration and cleaning with density-based spatial clustering of applications with noise (DBSCAN), a neural network verifies the labels with scatter of $\sim$14 K in \teff, 0.06 dex in \logg. There is a slight bias of 34 K with a scatter of 65 K between $ T_{\rm eff,slam}$ and $ T_{\rm eff,LAMOST}$. Lastly, we compared $T_{\rm eff,slam}$ with those of \citet{Mann-2015,Mann-2016}, who provided a metallicity-independent empirical relationship between \teff\ and color index from the Two Micron All Sky Survey \citep[2MASS,][]{Skrutskie-2006} and Gaia (see in Appendix).  
As shown in the right panels of Figure \ref{fig:SLAM_teff}, $\rm T_{eff,Mann}$ is systematically overestimated by 146 $\pm$ 45 K compared to $\rm T_{eff,slam}$.

\subsubsection{\logg\ comparison}
\begin{figure*}
\centering
\includegraphics[width=1.\textwidth, trim=0.cm 0.0cm 0.0cm 0cm, clip]{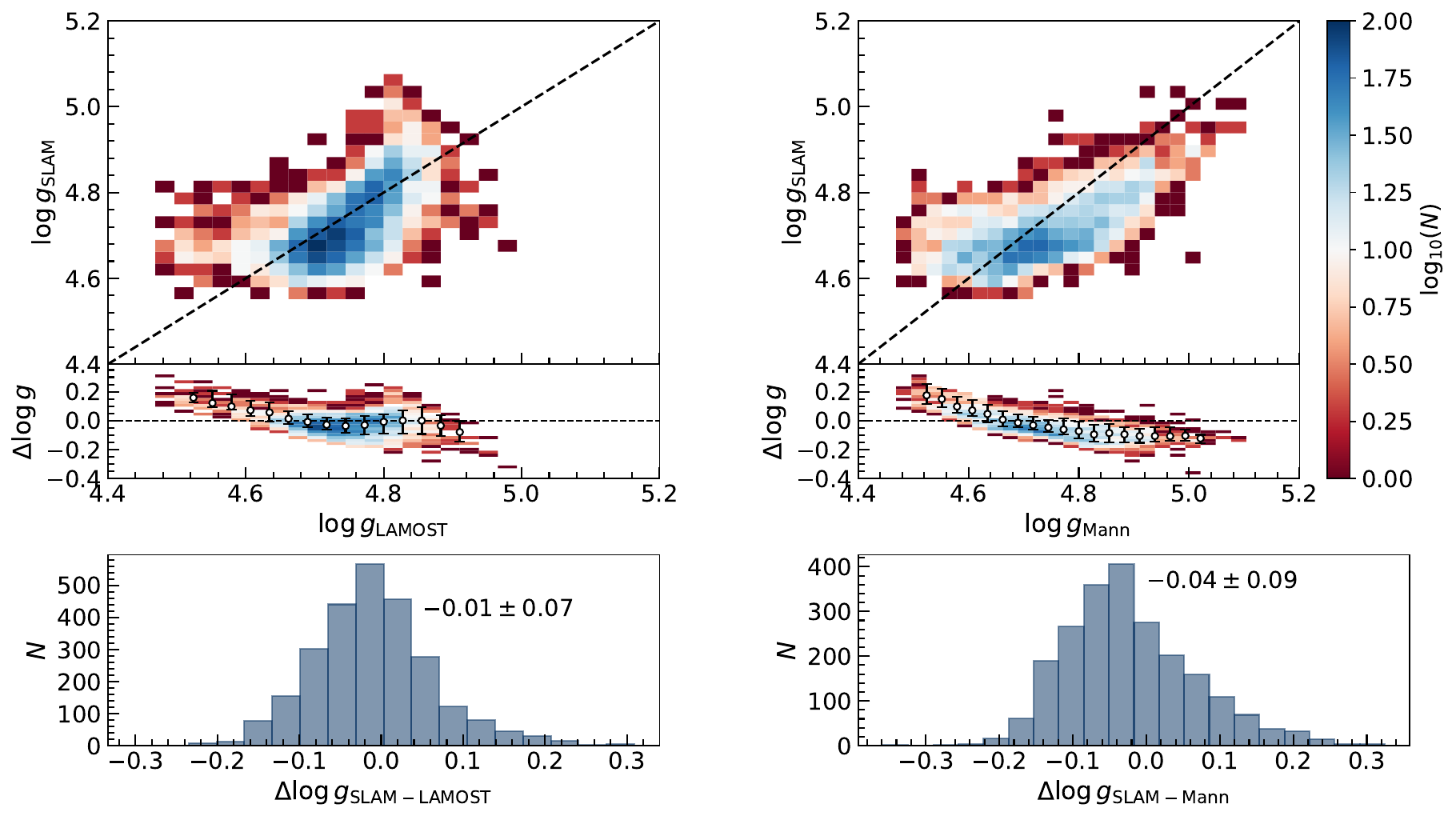}
\caption{Comparison of the SLAM-predicted surface gravity ($\log{g}{\rm slam}$) with (left) the LAMOST values ($\log{g}{\rm LAMOST}$) and (right) the values derived from \citet{Mann-2016,Mann-2019} ($\log{g}_{\rm Mann}$). The format is identical to that of Figure \ref{fig:SLAM_teff}. }\label{fig:SLAM_logg}
\end{figure*}

For the surface gravity, we compared $\log{g}_{\rm slam}$ with those of LAMOST gM/dM/sdM stars catalog \citep[$\log{g}_{\rm LAMOST}$;][]{Du-2024ApJS..275...42D}, as shown in the left panels of Figure \ref{fig:SLAM_logg}. The results demonstrate an excellent agreement, with no bias of -0.01 $\pm$ 0.07 dex. Additionally, the surface gravity of M dwarfs was determined by employing the following relation:
\begin{equation}\label{eq:logg}
\log {g} = 4.438 + \log_{10}(M_{*}/M_{\odot}) - 2 \log_{10}(R_{*}/R_{\odot})
\end{equation}
Here, $R_{\odot}$ and $M_{\odot}$ denote the solar radius and solar mass, these adopted values are taken from \citet{Mann-2016} and \citet{Mann-2019}, respectively. Further details are provided in the Appendix. 
 The right panels of Figure \ref{fig:SLAM_logg} display the comparison results of $\log{g}_{slam}$ and $\log{g}_{\rm Mann}$. It demonstrates that the $\log{g}_{\rm Mann}$ is consistent with $\log{g}_{\rm slam}$ within 0.04$\pm$0.09 dex. The residual of surface gravity ($\Delta \log{g}=\log{g}_{\rm SLAM}-\log{g}_{\rm Mann}$) show a slight systematic trend that our method yields higher $\log{g}_{\rm SLAM}$ values for stars with lower $\log{g}_{\rm Mann}$ ($<$4.7 dex) and lower $\log{g}_{\rm SLAM}$ values for stars with higher $\log{g}_{\rm Mann}$. This indicates a parameter-dependent discrepancy between these two methods, likely stemming from differences in analysis techniques. Considering the inherent difficulty in measuring $\log{g}$ for these cool stars, the level of disagreement shown here is within an acceptable range.

All above comparison results are shown in Table \ref{table:comparison}. It demonstrate that the stellar parameters derived from the SLAM model are consistent with certain other methods used to derive these parameters.

\begin{table}[ht]
\centering
\caption{Discrepancies between SLAM parameters and literature values.}\label{table:comparison}
\begin{tabular}{lccl}
\hline
Literature & \feh (dex) & \teff (K) &\logg (dex)  \\ \hline
test dataset&0.03$\pm$0.25&11$\pm168$&0$\pm$0.10\\
M+M wide binaries & -0.01$\pm$0.11 &   &  \\
\citet{Birky-2020} & -0.06$\pm$0.16 & -27$\pm$92  &  \\
\citet{Behmard-2025arXiv250114955B} &0.02$\pm$0.14&   & \\
\citet{Du-2024ApJS..275...42D} &  & -34$\pm$65&-0.01$\pm$0.07 \\
\citet{Mann-2015,Mann-2016}&  & -146
$\pm$45& \\
\citet{Mann-2016,Mann-2019}&  & & -0.04$\pm$0.09\\
\hline
\end{tabular}
\end{table}

\subsubsection{Distributions of SLAM parameters}
\begin{figure*}
\centering
\includegraphics[width=0.9\textwidth, trim=0.cm 0.0cm 0.0cm 0cm, clip]{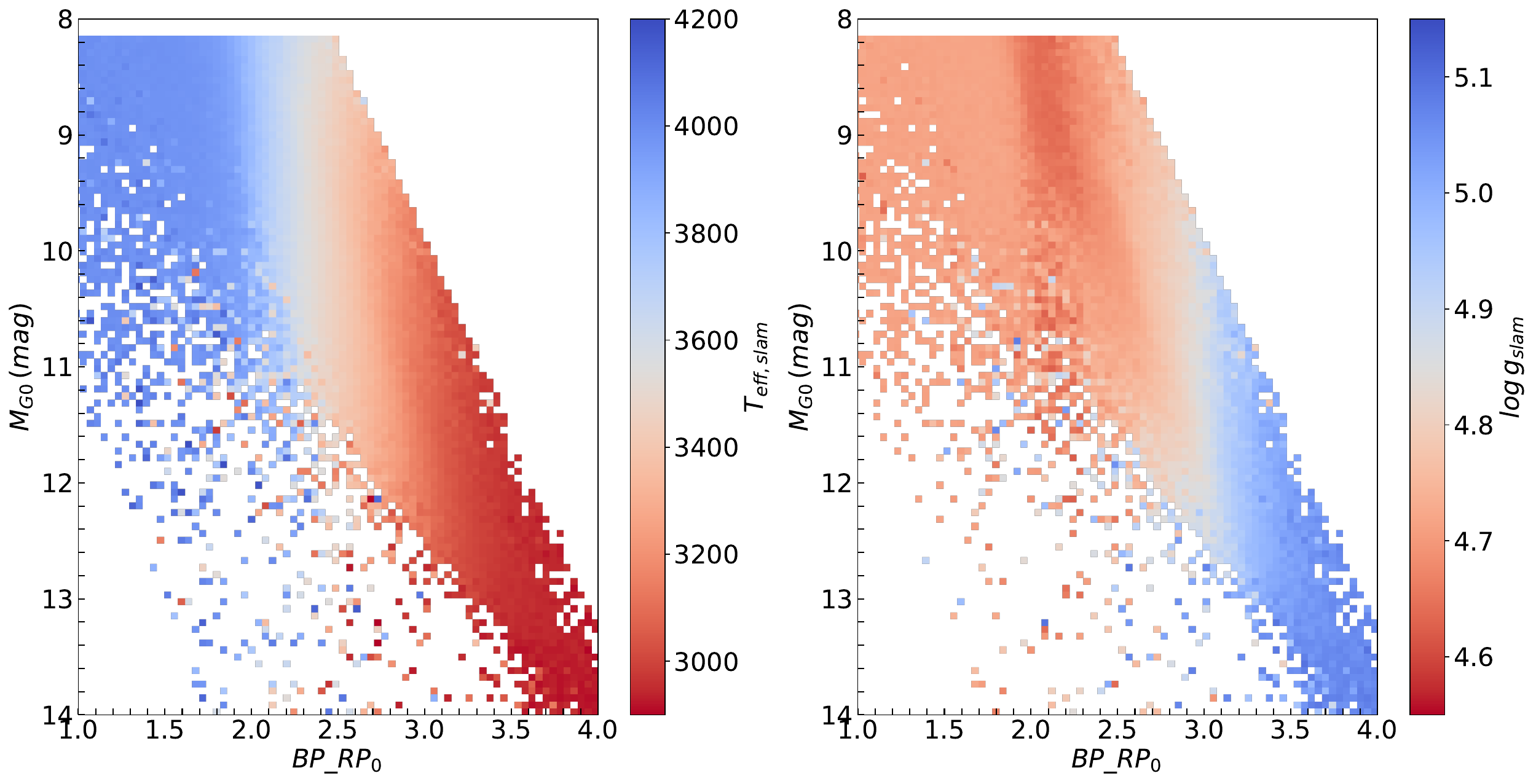}
\caption{The CMDs of all BOSS DR19 M dwarfs. The left panel is color-coded by the median values of $ T_{\rm eff,slam}$ of M dwarfs in each $BP\_RP_{0}$ and $M_{G0}$ bin, while the right panel is color-coded by the $\log{g}_{\rm slam}$.}\label{fig:teff_parsec}
\end{figure*}

In Figure \ref{fig:teff_parsec}, we present the distributions in the color-magnitude diagram of \teff\ (left) and \logg\ (right) derived from the SLAM model for all BOSS M dwarfs identified in Section \ref{sect:Identification of M dwarfs}. The distribution of \feh\ refer to Section \ref{sect:cali_ASPCAP}. Two panels of Figure \ref{fig:teff_parsec} reveal a clear correlation between effective temperature or surface gravity and the color or absolute magnitude index, which is expected given the intrinsic properties of M dwarfs.   
These results also demonstrate the reliability of the atmospheric parameters derived from the SLAM model. 

\section{Calibration of ASPCAP metallicity}\label{sect:cali_ASPCAP}

We cross-matched our M dwarf sample with APOGEE ASPCAP DR19. We applied quality cuts
to ensure reliable metallicity measurements by requiring a $\mathrm SNR$ greater than 20 for APOGEE spectra and 10 for BOSS spectra, respectively. Additionally, we restricted the sample to $\rm [Fe/H]_{slam}> -0.6$ dex as the SLAM model is not suitable for extrapolation, and most of our training data fall within this metallicity range.

\begin{figure*}
\centering
\includegraphics[width=1\textwidth, trim=0.cm 0.0cm 0.0cm 0cm, clip]{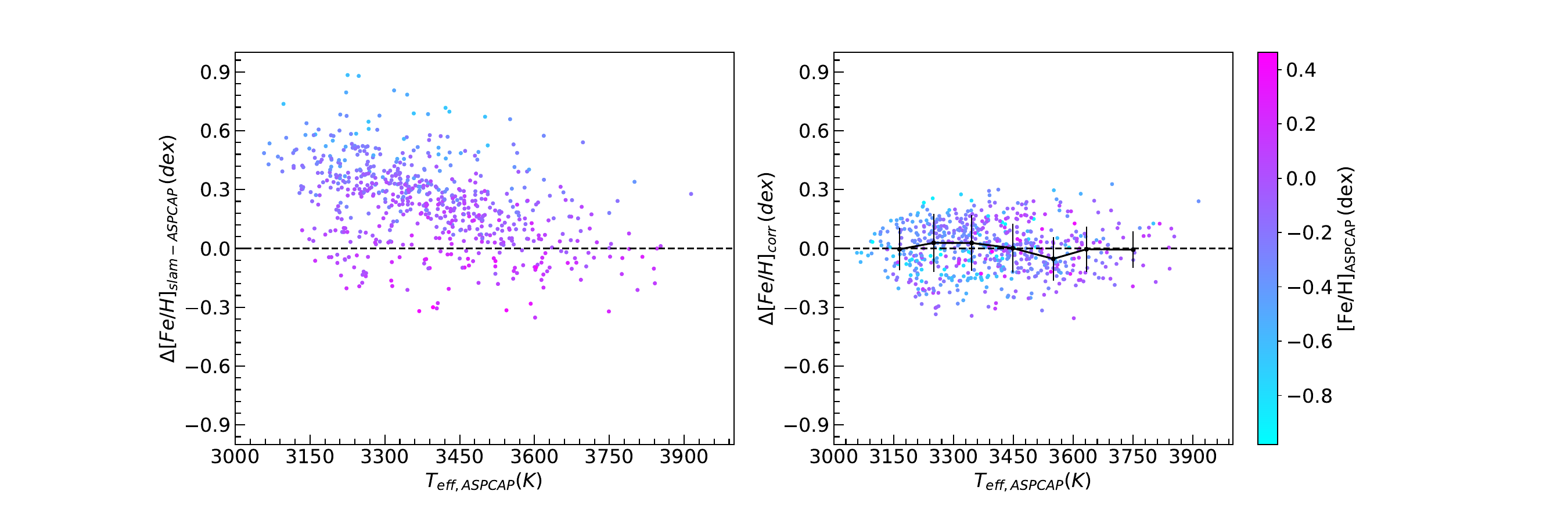}
\caption{The left panel displays the distribution of temperature of ASPCAP DR19 ($\rm T_{eff,ASPCAP}$) and the metallicity difference between SLAM prediction and that of ASPCAP DR19 ($\rm \Delta[Fe/H]=[Fe/H]_{slam}-[Fe/H]_{ASPCAP}$) of M dwarfs. The right panel is the same as the left panel, but the y-axis is the $\rm \Delta[Fe/H]_{corr}=[Fe/H]_{slam}-([Fe/H]_{ASPCAP}+\Delta_{{cal}})$, where the $\Delta_{\mathrm{cal}}$ refers to Equation \ref{eq:AP_cal}. Two panels are color-coded by the metallicity of ASPCAP.}\label{fig:ASPCAP_calib}
\end{figure*}

We compared the SLAM-derived metallicities ($\rm [Fe/H]_{slam}$) with those of ASPCAP DR19 ($\rm [Fe/H]_{ASPCAP}$). The results revealed that the difference between $\rm [Fe/H]_{slam}$ and $\rm [Fe/H]_{ASPCAP}$ depends not only on $\rm [Fe/H]_{ASPCAP}$ but also on the effective temperature of ASPCAP ($T_{\rm eff,ASPCAP}$), as shown in the left panel of Figure \ref{fig:ASPCAP_calib}. 
To address this systematic offset, we developed a calibration function based on $T_{\rm eff,ASPCAP}$ and $\rm [Fe/H]_{ASPCAP}$ to correct the ASPCAP metallicities. The calibration function is:

\begin{equation}\label{eq:AP_cal}
\begin{split}
\rm \Delta [Fe/H]_{cal}=& -0.427 x^{2}-0.158 y^{2}+0.431 xy\\
& +2.767 x-2.156 y-4.405
\end{split}
\end{equation}
where $x$ and $y$ are $T_{\rm eff,ASPCAP}/1000$ and $\rm [Fe/H]_{ASPCAP}$, respectively. The right panel  shows that the difference between $\rm [Fe/H]_{slam}$ and the corrected $\rm [Fe/H]_{ASPCAP}$ is close to zero across most of the temperature range.

\begin{figure*}[htbp!]
\centering
\includegraphics[width=0.95\textwidth, trim=0.cm 0.0cm 0.0cm 0cm, clip]{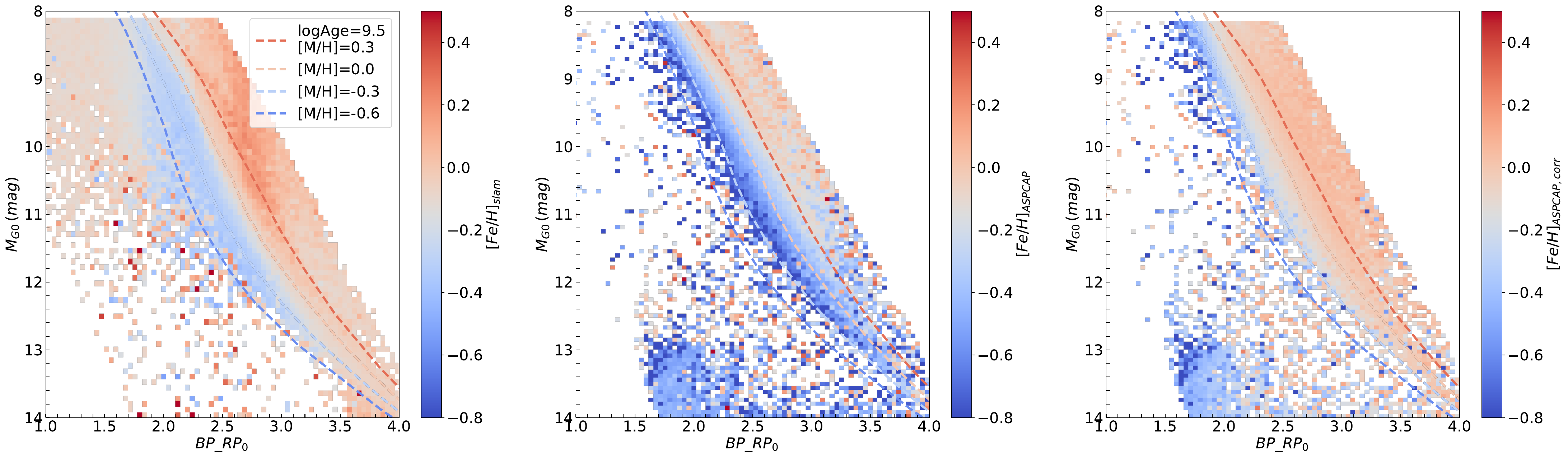}
\caption{The left panel displays the distribution of $BP\_RP_{0}$ and $M_{G0}$ for all BOSS M dwarfs, the colors are the median values of $\rm [Fe/H]_{slam}$ in each $BP\_RP_{0}$ and $M_{G0}$ bin. The middle panel presents the $BP\_RP_{0}$ versus $M_{G0}$ of all APOGEE M dwarfs, the color-coded by the median values of $\rm [Fe/H]_{ASPCAP}$ in each $BP\_RP_{0}$ versus $M_{G0}$ bin. The right panel is the same as the middle one but color-coded by the corrected metallicity, that is $\rm [Fe/H]_{ASPCAP}+\Delta [Fe/H]_{cal}$, where $\rm \Delta [Fe/H]_{cal}$ is shown in Equation \ref{eq:AP_cal}. Four lines in each panel are from PARSEC with [M/H]=0.3 , 0 , -0.3 , and -0.6 dex at $\log$(Age)= 9.5, the colors refer to the color map.}\label{fig:PARSEC}
\end{figure*}

We evaluated the metallicities of SLAM, ASPCAP, and the corrected ASPCAP by comparing them with the PAdova and TRieste Stellar Evolution Code \citep[PARSEC;][]{Bressan-2012}, as shown in Figure \ref{fig:PARSEC}. The lines in each panel display the isochrones with $\log$(Age) = 9.5 at different metallicities: [M/H] = 0.3 , 0.0 , -0.3 , and -0.6 dex. 

The left panel demonstrates that the metallicity estimated from the SLAM are in good agreement with the PARSEC model, particularly for stars with $\rm [Fe/H]_{slam} > -0.3$ dex. However, since the SLAM model is not designed for extrapolation and the training dataset contains fewer than 20 stars with \feh $<$ -0.6 dex, its predictions for metal-poor stars come with large uncertainties. This limitation explains the discrepancy between our metallicity distribution and the PARSEC model at lower metallicities (\feh $<$ -0.6 dex). In the middle panel, it demonstrates that the ASPCAP's 
metallicity is systematically underestimated compared to the PARSEC model. This finding is consistent with previous studies \citep{Qiu-2024MNRAS.52711866Q,Souto-2022ApJ...927..123S}, 
which have reported an underestimation of ASPCAP metallicities by approximately 0.10–0.24 dex compared to the metallicity of M dwarfs calibrated by using F, G, or K dwarf companions. In contrast, the right panel illustrates that the corrected ASPCAP metallicities, $\rm [Fe/H]_{ASPCAP,corr}$, align well with the PARSEC model. Notably, the distribution of $\rm [Fe/H]_{ASPCAP,corr}$ at lower metallicities shows better agreement with the PARSEC model compared to that of the $\rm [Fe/H]_{slam}$. It indicates that the correction effectively mitigates the systematic bias in ASPCAP's metallicity estimates.

\section{Conclusion and Discussion}\label{sect:conclu}
We developed a SLAM model for SDSS-V/BOSS M dwarfs, which can derive the atmospheric parameters (\feh, \teff, and \logg) of M dwarfs from optical low-resolution spectra. The metallicity (\feh) was calibrated from LAMOST F, G, and K dwarf companions, while the effective temperature (\teff) and surface gravity (\logg) were calibrated by the APOGEE Net pipeline. The uncertainties of the SLAM predictions for \feh, \teff, and \logg\, depend on the spectral $\mathrm {SNR}$, reaching 0.19 dex, 132 K, and 0.10 dex, respectively, at a $\mathrm {SNR}$ of 10. 

For the test dataset, the SLAM output parameters show excellent agreement with the reference values. The median and standard deviation of the differences for \feh, \teff, and \logg\, are 0.03  $\pm$ 0.25 dex, 11 $\pm$ 168 K, and 0 $\pm$ 0.10 dex, respectively.

We compared the SLAM-predicted \feh\ values between the primary and secondary M dwarfs in wide binary systems and found no significant bias (0.01 dex), with a scatter of 0.11 dex. Additionally, we compared the SLAM predicted \feh\  with those from other two studies: \citet{Birky-2020} and \citet{Behmard-2025arXiv250114955B}. Both studies also calibrated the \feh\ of M dwarfs using F, G, and K dwarf companions. The differences between our results and those of \citet{Birky-2020} and \citet{Behmard-2025arXiv250114955B} are 0.06 $\pm$ 0.16 dex and 0.02 $\pm$ 0.14 dex, respectively. 

For the effective temperature, our results agree well with those of \citet{Birky-2020}, showing a bias of -27 $\pm$ 92 K. Compared to the \teff\ values from LAMOST, there is a systematic difference of -34 ± 65 K. We also compared $\rm T_{eff,slam}$ with those from \citet{Mann-2015,Mann-2016}, who determined \teff\ using an empirical relationship between temperature and magnitude. The results indicate that the \teff\ values from \citet{Mann-2015,Mann-2016} are systematically overestimated by 146 $\pm$ 45 K.

For the surface gravity ($\log{g}_{\rm slam}$), we compared it with that from LAMOST, finding excellent agreement with no bias of -0.01 ± 0.06 dex. Additionally, we determined the surface gravity of M dwarfs using stellar radii and masses according to \citet{Mann-2015,Mann-2016,Mann-2019}. The difference between $\log{g}_{\rm slam}$ and their values is -0.04 $\pm$ 0.09 dex.

These comprehensive comparisons demonstrate the reliability of stellar parameters for BOSS M dwarfs derived from the SLAM model. Our model successfully addresses the long-standing challenge in determining accurate parameters for M dwarf stars, particularly in metallicity measurements. The SLAM model is integrated into the Astra package and has been applied to all the BOSS M dwarfs of SDSS-V. The SLAM output parameters and their descriptions are listed in Table \ref{table:para_describe}, with detailed values available in the official dataset released by SDSS DR19.

A key limitation of the SLAM model is that it can not be used for extrapolation. Its predictions are trustworthy only within its well-defined training domain: \feh = [-0.6, 0.5] dex, \teff = [3100, 3900] K, and \logg = [4.45, 4.95] dex. Outside this domain, the scarcity of training data (N $<$ 20) renders the predictions unreliable with potentially large uncertainties.

Additonally, we found a bias in \feh\ between our work and ASPCAP, which shows a correlation with both ASPCAP's \feh\ and \teff. To address this, we developed a calibration model as a function of ASPCAP's \feh\ and \teff, with a detailed model presented in Subsection \ref{sect:cali_ASPCAP}.

The current training sample contains relatively few very metal-poor M dwarfs. At present we cannot yet determine whether this deficit is mainly caused by survey selection effects or reflects genuine astrophysical scarcity, and disentangling these possibilities is beyond the scope of this work. Future progress will require extending the low-metallicity training and validation set. In particular, we will search for additional FGK+M wide binaries in forthcoming SDSS releases (e.g., DR20), where the M-dwarf metallicity can be anchored to the well-measured [Fe/H] of the FGK primary. We will also identify metal-poor M-dwarf candidates through two complementary routes: halo-like kinematics/orbital-parameter selection and spectral template matching in BOSS. For candidates lacking FGK companions, high-resolution optical/NIR follow-up spectroscopy will provide independent labels. These steps will allow us to extend SLAM to lower metallicities and enable a dedicated study of metal-poor M dwarfs in future work.

\section*{acknowledgements}
D.S. thanks the National Council for Scientific and Technological Development – CNPq process No. 404056/2021-0. Y.Y.S. acknowledges support from the Dunlap Institute, which is funded through an endowment established by the David Dunlap family and the University of Toronto. R. L-V. acknowledges support from Secretar\'ia de Ciencia, Humanidades, Tecnolog\'ia e Inovacci\'on (SECIHTI) through a postdoctoral fellowship within the program ``Estancias posdoctorales por M\'exico''.  S. M. has been supported by the LP2021-9 Lend\"ulet grant of the Hungarian Academy of Sciences and by the NKFIH excellence grant TKP2021-NKTA-64.
Funding for the Sloan Digital Sky Survey V has been provided by the Alfred P. Sloan Foundation, the Heising-Simons Foundation, the National Science Foundation, and the Participating Institutions. SDSS acknowledges support and resources from the Center for High-Performance Computing at the University of Utah. SDSS telescopes are located at Apache Point Observatory, funded by the Astrophysical Research Consortium and operated by New Mexico State University, and at Las Campanas Observatory, operated by the Carnegie Institution for Science. The SDSS web site is \url{www.sdss.org}.

SDSS is managed by the Astrophysical Research Consortium for the Participating Institutions of the SDSS Collaboration, including Caltech, The Carnegie Institution for Science, Chilean National Time Allocation Committee (CNTAC) ratified researchers, The Flatiron Institute, the Gotham Participation Group, Harvard University, Heidelberg University, The Johns Hopkins University, L'Ecole polytechnique f\'{e}d\'{e}rale de Lausanne (EPFL), Leibniz-Institut f\"{u}r Astrophysik Potsdam (AIP), Max-Planck-Institut f\"{u}r Astronomie (MPIA Heidelberg), Max-Planck-Institut f\"{u}r Extraterrestrische Physik (MPE), Nanjing University, National Astronomical Observatories of China (NAOC), New Mexico State University, The Ohio State University, Pennsylvania State University, Smithsonian Astrophysical Observatory, Space Telescope Science Institute (STScI), the Stellar Astrophysics Participation Group, Universidad Nacional Aut\'{o}noma de M\'{e}xico, University of Arizona, University of Colorado Boulder, University of Illinois at Urbana-Champaign, University of Toronto, University of Utah, University of Virginia, Yale University, and Yunnan University.
Guoshoujing Telescope (the Large Sky Area Multi-Object Fiber Spectroscopic Telescope LAMOST) is a National Major Scientific Project built by the Chinese Academy of Sciences. Funding
for the project has been provided by the National Development
and Reform Commission. LAMOST is operated and managed by
the National Astronomical Observatories, Chinese Academy of Sciences.

\bibliography{sample631}
\bibliographystyle{aasjournal}

\section*{Appendix}\label{sec:appendix}
\begin{figure*}
\centering
\includegraphics[width=1\textwidth, trim=0.cm 0.0cm 0.0cm 0cm, clip]{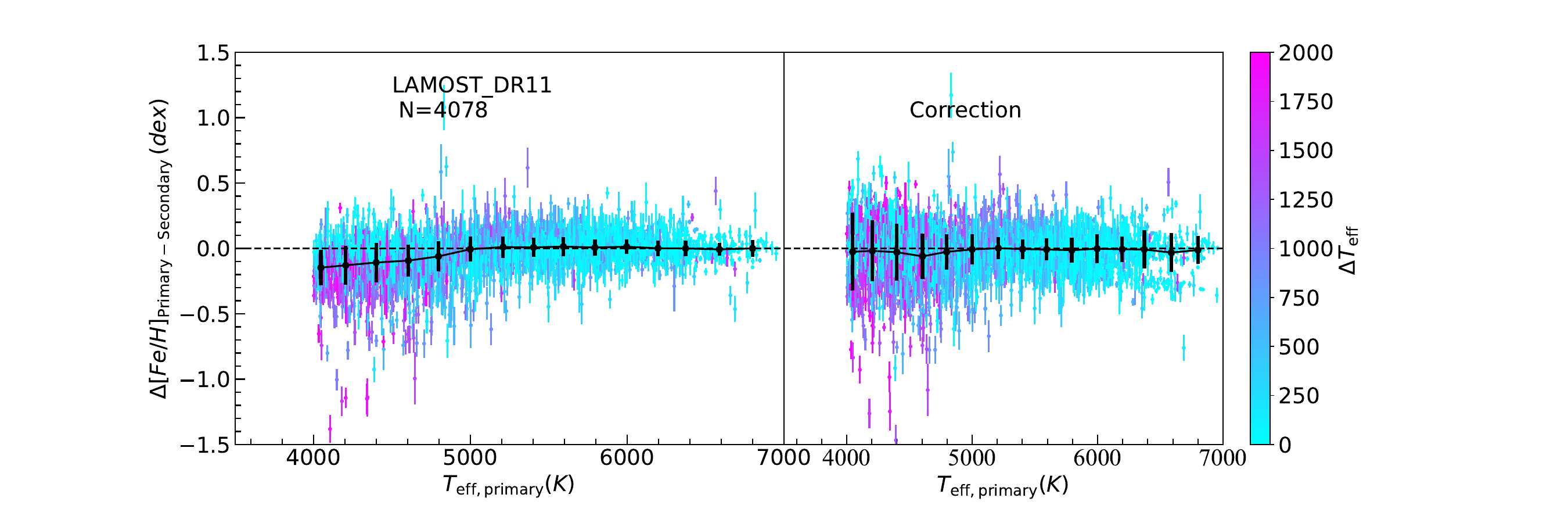}
\caption{4078 FGK+FGK wide binaries from LAMOST DR11. In the left panel, the x-axis is the \teff\ of the primary stars, the y-axis is the metallicity difference between the primary ($\rm [Fe/H]_{primary}$) and secondary ($\rm [Fe/H]_{secondary}$) stars. The right panel is the same as the left one, but the y-axis is the difference in corrected \feh\ according to Equation (\ref{eq:deltafeh}).}\label{fig:FGK_FGK_feh}
\end{figure*}

\citet{Niu-2023ApJ...950..104N} selected 2,296 FGK+FGK dwarf wide binaries from the LAMOST DR7 A/F/G/K stars catalog to investigate the \feh\ of both components in each binary system. They found that the \feh\ values in the A/F/G/K stars catalog require a temperature-based adjustment. To address this, they developed a temperature-dependent broken power-law model to calibrate the \feh\ of dwarfs with 4000 $<$ \teff\ $<$ 7000 K in the A/F/G/K stars catalog. The calibration model is as follows:

\begin{equation}
 \Delta [Fe/H] = \left\{
    \begin{aligned} 
    0.358 *(T_{\rm eff} /5281.4) ^{-2.404}-0.4 &,\, 5281.4 \leq T_{\rm eff}  \\
    0.358 *(T_{\rm eff} /5281.4) ^{1.254}-0.4 &,\,\, 5281.4> T_{\rm eff}.
    \end{aligned}
 \right.
 \label{eq:deltafeh}
\end{equation}

where \teff\ is from the LAMOST. Following the methodology of \citet{Niu-2023ApJ...950..104N}, we identified 4078 FGK+FGK wide binaries from LAMOST DR11. The \feh\ difference between the primary ($\rm [Fe/H]_{primary}$) and secondary ($\rm [Fe/H]_{secondary}$) stars versus \teff\ of primary stars is displayed in the left panel of Figure \ref{fig:FGK_FGK_feh}. We applied the model of \citet{Niu-2023ApJ...950..104N} to calibrate the \feh\ of primary and secondary stars, the difference in corrected \feh\ versus the temperature of the primary stars as shown in the right panel. The corrected $\rm \Delta [Fe/H]=([Fe/H]_{primary}-[Fe/H]_{secondary})$ becomes independent of \teff\ and converges to zero. We added the corrected \feh\ of LAMOST F, G and K dwarfs as one of the training labels of SLAM model. 

In addition to \teff, \logg, and \feh, we also incorporated \alp\ as a training label in the SLAM model, with the values sourced from ASPCAP. The scarcity of external validation data for \alp of M dwarfs, however, prevents a further discussion of its performance.

The $T_{\rm eff,mann}$ and $\log{g}_{\rm mann}$ used in Subection \ref{sect:Validation} derived from \citet{Mann-2015,Mann-2016} and \citet{Mann-2016,Mann-2019} respectively. For $T_{\rm eff,mann}$, they provided a metallicity-independent empirical relationship between \teff\ and magnitudes from the Two Micron All Sky Survey \citep[2MASS, $J, H$ bands;][]{Skrutskie-2006} and Gaia ($BP , RP$ bands). The formula is:

\begin{equation}\label{eq:teff_mann}
T_{\rm eff}/3500 =a+bX+cX^{2}+dX^{2}+eX^{2}+f(J-H)+g(J-H)^{2} 
\end{equation}
where $X$ is the color index $BP-RP$. We used the coefficients for the $BP-RP$, $J-H$ relation from Table 2 in \citet{Mann-2016}, which has a scatter of 49 K. 
That is, $a=3.172$, $b=-2.475$, $c=1.082$, $d=-0.2231$, $e=0.01738$, $f=0.08776$, and $g=0.04355$.
For the determination of $\log{g}_{\rm mann}$, the $R_{\odot}$ and $M_{\odot}$ used in equation (\ref{eq:logg}) were inferred by adopting the relationship between $R_{*}/R_{\odot}$ and absolute magnitude in 2MASS $Ks$ band, as described in Table 1 of \citet{Mann-2016}, which shows a $\sigma$\, of 2.89\%. That is,
\begin{equation}\label{eq:R}
R_{*} = 1.9515-0.3520 M_{Ks}+(0.0168)^{2}.
\end{equation}
The stellar mass was determined by the relationship between $M_{*}/M_{\odot}$ and the absolute magnitude in 2MASS $Ks$ band provided in \citet{Mann-2019}. In particular, we used the one that shows $n$=5 and BIC of 86 in Table 6 recommended by \citet{Mann-2019} to derive $M_{*}/M_{\odot}$ of M dwarfs in this work. i.e.,
\begin{equation}\label{eq:mass}
\log_{10}(\frac{M_{*}}{M_{\odot}})=\sum_{i=0}^{n} a_{i}(M_{Ks}-zp)^{i}. 
\end{equation}
where $zp$=7.5, $n$=5, $a_{0...5}$=-0.642, -0.208, $-8.43\times10^{-4}$,  $7.87\times10^{-3}$, $1.42\times10^{-4}$, $-2.13\times10^{-4}$.
Then $\log{g}_{\rm mann}$ can be determined according to equation (\ref{eq:logg}).

\begin{table*}
\centering
\caption{Catalog description of FGK+M wide binaries.\label{table:binary}}
\begin{tabular}{l l l}
\hline\hline
Column & Units & Description \\
\hline
\texttt{source\_id1/2} &  & Gaia source ID of LAMOST FGK primary / BOSS M dwarf secondary \\
\texttt{ra1/2}         & degree & Right ascension of primary/secondary stars \\
\texttt{dec1/2}        & degree & Declination of primary/secondary stars \\
\texttt{FeH}           & dex    & [Fe/H] of the wide binaries from LAMOST FGK dwarfs \\
\texttt{FeH\_niu}      & dex    & [Fe/H] calibrated following \citet{Niu-2023ApJ...950..104N} \\
\texttt{e\_FeH}        & dex    & Uncertainty in [Fe/H] \\
\texttt{Teff}          & K      & Effective temperature of M dwarfs from APOGEE Net \\
\texttt{e\_Teff}       & K      & Uncertainty in $T_{\rm eff}$ \\
\texttt{alphaM}        & dex    & [$\alpha$/M] from ASPCAP \\
\texttt{e\_alphaM}     & dex    & Uncertainty in [$\alpha$/M] \\
\texttt{logg}          & dex    & Surface gravity from APOGEE Net \\
\texttt{e\_logg}       & dex    & Uncertainty in $\log g$ \\
\texttt{SNR}           &        & Signal-to-noise ratio of the BOSS spectra \\
\hline
\end{tabular}

\medskip
\parbox{0.98\textwidth}{\footnotesize
This table presents the parameters of the FGK+M wide-binary systems selected in Section~\ref{sect:binary}.
The complete catalog is available at: \url{https://nadc.china-vo.org/res/r101717/LAMOST_FGK_BOSS_M_dwarfs.fits}.}
\end{table*}

\begin{table*}
\centering
\caption{Catalog description of SLAM output parameters.\label{table:para_describe}}
\begin{tabular}{l l l}
\hline\hline
Column & Units & Description \\
\hline
\texttt{Teff}         & K   & Effective temperature from SLAM \\
\texttt{e\_Teff}      & K   & Uncertainty in $T_{\rm eff}$ from SLAM \\
\texttt{logg}         & dex & Surface gravity from SLAM \\
\texttt{e\_logg}      & dex & Uncertainty in $\log g$ from SLAM \\
\texttt{alphaM}       & dex & [$\alpha$/M] from SLAM \\
\texttt{e\_alphaM}    & dex & Uncertainty in [$\alpha$/M] from SLAM \\
\texttt{FeH}          & dex & [Fe/H] from SLAM, calibrated using LAMOST FGK companions \\
\texttt{e\_FeH}       & dex & Uncertainty in [Fe/H] from SLAM \\
\texttt{FeH\_niu}     & dex & [Fe/H] calibrated following \citet{Niu-2023ApJ...950..104N} \\
\texttt{e\_FeH\_niu}  & dex & Uncertainty in [Fe/H]\_niu from SLAM \\
\hline
\end{tabular}

\medskip
\parbox{0.98\textwidth}{\footnotesize
The SLAM parameters are released in SDSS DR19 at:
\url{https://data.sdss.org/sas/dr19/env/MWM_ASTRA/0.6.0/summary/astraAllStarSlam-0.6.0.fits.gz}.}
\end{table*}



\end{document}